%% file: anica.tex
\documentclass[acmsmall,screen]{acmart}

\setcopyright{rightsretained}
\acmPrice{}
\acmDOI{10.1145/3563288}
\acmYear{2022}
\copyrightyear{2022}
\acmSubmissionID{oopslab22main-p75-p}
\acmJournal{PACMPL}
\acmVolume{6}
\acmNumber{OOPSLA2}
\acmArticle{125}
\acmMonth{10}

\usepackage{booktabs}   
\usepackage{subcaption} 
\usepackage{microtype}

\usepackage[english]{babel}
\addto\extrasenglish{%
}

\usepackage[fleqn]{mathtools}
\usepackage[linesnumbered,noend]{algorithm2e}
\usepackage{listings}
\usepackage{csquotes}
\usepackage{tabularx}
\usepackage{multirow}
\usepackage{enumitem}  

\usepackage{tikz}
\usetikzlibrary{positioning}
\usetikzlibrary{arrows}
\usetikzlibrary{calc}

\usepackage{standalone}

\usepackage{pgfplotstable}

\usepackage{todonotes}

\usepackage{adjustbox}
\usepackage{array}

\newcolumntype{R}[2]{%
    >{\adjustbox{angle=#1,lap=\width-(#2)}\bgroup}%
    l%
    <{\egroup}%
}
\newcommand*\rot{\multicolumn{1}{R{32}{1em}}}
\newcommand*\rotless{\multicolumn{1}{R{30}{1em}}}

\SetAlgorithmName{Alg.\!}{Algorithm}{Algorithm}
\SetAlgoCaptionSeparator{.}

\SetAlCapSkip{0.6em}
\SetKw{Continue}{continue}

\SetCommentSty{mycommfont}

\input{tex/definitions}

\begin{document}

\title{AnICA: Analyzing Inconsistencies in Microarchitectural Code Analyzers}         


\author{Fabian Ritter}
\orcid{0000-0001-9227-0910}
\affiliation{
  \institution{Saarland University}
  \streetaddress{Saarland Informatics Campus}
  \city{Saarbrücken}
  \country{Germany}
}
\email{fabian.ritter@cs.uni-saarland.de}

\author{Sebastian Hack}
\orcid{0000-0002-3387-2134}
\affiliation{
  \institution{Saarland University}
  \streetaddress{Saarland Informatics Campus}
  \city{Saarbrücken}
  \country{Germany}
}
\email{hack@cs.uni-saarland.de}

\begin{abstract}
\input{tex/abstract}

\end{abstract}



\maketitle

\input{tex/intro}

\input{tex/background}

\input{tex/main}

\input{tex/eval}

\input{tex/relwork}

\input{tex/conclusion}


\bibliography{references}

%

\end{document}

%% file: tex/definitions.tex
\usepackage{xspace}

\makeatletter
\newcommand\newtag[2]{#1\def\@currentlabel{#1}\label{#2}}
\makeatother

\newcommand{\llvm}{LLVM\xspace}
\newcommand{\llvmmca}{llvm-mca\xspace}
\newcommand{\exegesis}{EXEgesis\xspace}

\newcommand{\uopsinfo}{uops.info\xspace}
\newcommand{\nanobench}{nanoBench\xspace}
\newcommand{\uica}{uiCA\xspace}

\newcommand{\iaca}{IACA\xspace}

\newcommand{\cqa}{CQA\xspace}

\newcommand{\pmevo}{PMEvo\xspace}
\newcommand{\palmed}{Palmed\xspace}

\newcommand{\osaca}{OSACA\xspace}

\newcommand{\ithemal}{Ithemal\xspace}
\newcommand{\difftune}{DiffTune\xspace}
\newcommand{\bhive}{BHive\xspace}

\newcommand{\bigland}{\bigwedge}

\newcommand{\nat}{\ensuremath{\mathbb{N}}\xspace}

\def\eg{e.g.\@\xspace}
\def\ie{i.e.\@\xspace}
\def\cf{cf.\@\xspace}

\newcommand{\approach}{AnICA\@\xspace}

\newcommand{\concdomain}{\mathbb{C}}
\newcommand{\absdomain}{\mathbb{A}}
\newcommand{\bbs}{\mathbb{B}}
\newcommand{\insns}{\mathsf{Insns}}

\newcommand{\set}[1]{\bigl\{ #1 \bigr\}}
\newcommand{\given}{\mathrel{\bigm\vert}}

\newcommand{\powset}[1]{\mathcal{P}(#1)}
\newcommand{\fundom}[1]{\mathsf{dom}(#1)}

\newcommand{\definedas}{\mathrel{:=}}
\newcommand{\definedeq}{\mathrel{:\Leftrightarrow}}

\newcommand{\uops}{$\mu$ops\xspace}

\providecommand{\abs}[1]{\lvert#1\rvert}
\DeclarePairedDelimiter\floor{\lfloor}{\rfloor}

\newcommand{\idxalias}{\mathit{al}}
\newcommand{\idxinsn}{\mathit{in}}

\theoremstyle{definition}
\newtheorem{exmp}{Example}
\newtheorem*{defn}{Definition}

%% file: tex/abstract.tex
Microarchitectural code analyzers, \ie, tools that estimate the throughput of machine code basic blocks, are important utensils in the tool belt of performance engineers.
Recent tools like \llvmmca, \uica, and \ithemal use a variety of techniques and different models for their throughput predictions.
When put to the test, it is common to see these state-of-the-art tools give very different results.
These inconsistencies are either errors, or they point to different and rarely documented assumptions made by the tool designers.

In this paper, we present \approach, a tool taking inspiration from differential testing and abstract interpretation to systematically analyze inconsistencies among these code analyzers.
Our evaluation shows that \approach can summarize thousands of inconsistencies in a few dozen descriptions that directly lead to high-level insights into the different behavior of the tools.
In several case studies, we further demonstrate how \approach automatically finds and characterizes known and unknown bugs in \llvmmca, as well as a quirk in AMD's Zen microarchitectures.

%% file: tex/intro.tex
\section{Introduction}
\label{sec:intro}

Making software run faster has always been a major goal for programmers as well as for computer science research.
To make software run as fast as possible, we need to have an understanding of how fast some code will execute on a given machine.
Recently, research has seen a rise of interest in such an understanding at the lowest level: estimating the throughput of CPU-bound loop-free instruction sequences.
This is witnessed by the wide range of microarchitectural code analyzers that give such instruction throughput estimates, \eg, \llvmmca~\cite{llvmmca}, \uica~\cite{abel21uica}, \osaca~\cite{laukemann18osaca}, \iaca~\cite{iaca}, \cqa~\cite{rubial14cqa}, \ithemal~\cite{mendis19ithemal}, and \difftune~\cite{renda2020difftune}.
While these tools vary in the methods they employ -- compiler scheduling models, microbenchmarks, or machine learning -- they share the goal of estimating the throughput of basic blocks on a given processor.
\begin{figure}
  \centering \makebox[0pt]{\scalebox{0.79}{\input{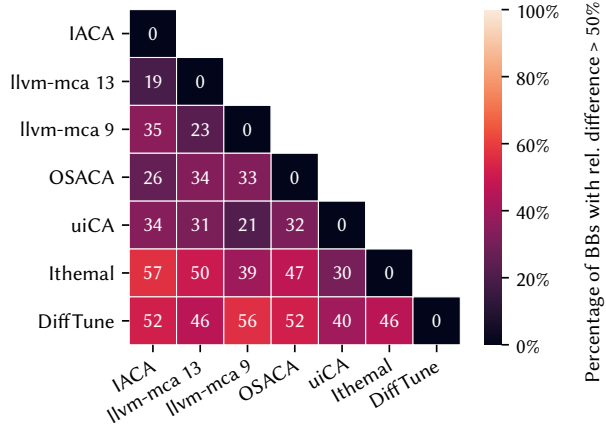}}}
  \caption{Heat map showing the percentage of basic blocks with throughput estimates that deviate by more than $50\%$ for each pair of predictors.}
  \label{fig:intro_motivation}
  \Description{
    The figure contains an entry for each pair made of the following throughput predictors: IACA, llvm-mca (version 13), llvm-mca (version 9), OSACA, uiCA, Ithemal, and Difftune.
    For every pairs, at least 19 percent of the basic blocks are predicted inconsistently (the lowest, at 19 percent, being IACA vs llvm-mca 13).
    Most pairs exceed 30 percent, and many of the pairs involving Ithemal and Difftune range between 40 and 57 percent (the highest, at 57 percent, being Ithemal vs IACA).
    The only exception is that all entries for comparing throughput a predictor with itself are 0.
  }
\end{figure}

However, these tools vary significantly in their predictions and deliver inconsistent results.
\autoref{fig:intro_motivation} shows the results of an experiment in which we randomly generated 10,000 basic blocks consisting of 4 instructions each\footnote{A basic block is generated by individually sampling instructions from the machine-readable x86-64 ISA description from \url{uops.info} and instantiating them with valid operands.} and let several throughput predictors give their estimate for these blocks assuming the Intel Haswell\footnote{Haswell is the only microarchitecture supported by all compared tools.} microarchitecture.
For each pair of throughput predictors, the heat map contains an entry indicating the percentage of basic blocks for which the throughput estimates deviated by more than $50\%$ of their average.

Overall, all pairs of predictors exhibit substantial numbers of inconsistencies.
As we can see from the inconsistencies in different versions of \llvmmca (23\% of the basic blocks in \autoref{fig:intro_motivation} are predicted inconsistently between \llvmmca versions 9 and 13), even closely related implementations are affected.
There may be several reasons for these deviations, \eg:
\begin{itemize}
  \item The different performance models might fail to capture relevant parts of the execution.
  \item The tools might be built with different (implicit or explicit) assumptions.
  \item The learning-based tools might need more training data.
  \item The implementations might contain bugs.
\end{itemize}

In any of these cases, finding and characterizing such inconsistencies is valuable.
If the cause is unintentional, they can help to improve the tools.
If the inconsistencies are the result of deliberate choices of the developers, identifying them helps us to explore the limits of the tools.
Therefore, the goal of this work is to automatically discover inconsistencies in the results of instruction throughput predictors and to give insight into their causes.

Our approach, \approach, applies differential testing~\cite{mckeeman98} to a pair of basic block throughput predictors.
The core idea is to randomly sample inputs and compare the outputs of the tools under investigation.
If the tools do not agree, we found an inconsistency.
Similar techniques are used in a variety of domains, prominently for compilers~\cite{mckeeman98}, SSL/TLS certificate validators~\cite{brubaker14frankencerts,chen15mucerts}, but also for software components close to the processor like instruction decoders~\cite{paleari2010,jay2018,woodruff2021differential}.

The existing approaches do however not transfer well to basic block throughput predictors since those provide an unusual setting for differential testing:
the above heat map shows that finding inputs that exhibit inconsistencies is not difficult; elaborate methods for searching the input space are therefore not necessary.
However, just listing the large amount of inconsistencies we find would also not be very helpful to understand and improve the tools under test.
The focus of \approach is therefore to find compact characterizations of large classes of inputs that cause inconsistencies.
We apply concepts from abstract interpretation~\cite{cousot77} to find these compact characterizations and present them together with witnesses for their derivation.
These witnesses give insights in two directions:
\begin{itemize}
  \item They contain examples of represented basic blocks that exhibit inconsistencies and
  \item they show the boundaries of the problem through similar basic blocks that do not exhibit inconsistencies.
\end{itemize}

For many of the combinations of tools shown in \autoref{fig:intro_motivation}, ten of \approach's discoveries are sufficient to characterize more than half of the several thousand encountered inconsistencies.
We investigate results of \approach in case studies showing that the results are helpful for improving performance models in several ways:
to find modeling bugs and regressions from one tool version to the next, to understand differing modeling assumptions, and to identify underrepresented constructs in the training sets of learned predictors.

In summary, we make the following contributions:
\begin{itemize}[leftmargin=*]
  \item A novel algorithm based on differential testing and using concepts from abstract interpretation to find compact characterizations of inputs that cause inconsistencies in basic block throughput predictors, presented in \autoref{sec:main}.
  \item A modular and extensible implementation of this algorithm targeting throughput predictors for the x86-64 instruction set architecture that we evaluate in \autoref{sec:eval}.
  \item Our case studies show how \approach exposes subtle modeling differences between the tools, identifies a long-standing crash in \llvmmca with a two-instruction test case, and characterizes a series of inaccuracies in \llvmmca's model for the AMD Zen+ microarchitecture.
    In this process, \approach even finds an unusual quirk in the Zen+ microarchitecture itself. (\autoref{sec:case-studies})
\end{itemize}

%% file: tex/background.tex
\section{Background: Predicting Throughput}
\label{sec:background}

There is a wide range of approaches that estimate the performance of programs.
They differ in the kind of input they expect -- from short sequences of machine instructions~\cite{iaca} to entire programs~\cite{binkert11gem5} -- as well as the accuracy they strive for -- from cycle-accurate~\cite{bohm10} to \enquote{back-of-the-envelope calculations}~\cite{williams09roofline,ofenbeck14roofline}.

This work analyzes tools at a specific point of this spectrum: low-level throughput predictors for \emph{basic blocks}, \ie, short sequences of machine instructions without control flow.
Throughput here means the sustained rate at which the basic block can be executed infinitely often, either as instructions/iterations per clock cycle or as number of cycles required to execute one instance.
These tools typically aim to be close to cycle-accurate, but cannot do an exact simulation of the target hardware.
Instead, they use a model of the hardware that is built either from proprietary knowledge about the hardware~\cite{iaca}, from the scheduling models of a compiler~\cite{llvmmca}, through micro-benchmarking~\cite{abel21uica}, or via machine learning~\cite{mendis19ithemal,renda2020difftune}.

Common assumptions for these tools are that all memory accesses hit the L1 cache and that execution of the basic block is in a steady state (it is the body of an innermost loop that is executed indefinitely often).
With these properties, the main domain of application for such throughput predictors is in the optimization of short, very hot code regions in programs where performance is crucial.

Even with these common assumptions, the task of predicting the throughput for a given basic block is not easy.
Modern processors split instructions into undocumented micro operations (\enquote{$\mu$ops}) and reorder them as freely as the data dependencies allow.
Numerous undocumented buffers and execution units make the processor fast, but they also impede accurate throughput estimation.

Moreover, there is often not a single well-defined throughput for a given basic block on a microarchitecture~\cite{abel21uica}:
\begin{itemize}[leftmargin=*]
  \item On many microarchitectures, the execution time of some instructions depends on their input values.
  \item Basic blocks might contain data dependencies if certain input values are pointers to the same memory location.
  \item Depending on whether the basic block is repeated through a loop or through concatenating many copies, different bottlenecks determine the throughput.
\end{itemize}
If a tool wants to predict the throughput of arbitrary basic blocks, it needs to assume behaviors for all such points.
Very often, basic block throughput predictors do not come with an explicit statement of these assumptions.
This makes it difficult to judge which tool is better suited for which task:
If two tools are based on different assumptions, we cannot just compare their accuracy with respect to a common ground truth since such a ground truth needs to depend on the chosen assumptions.

In the following section we describe how \approach compares basic block throughput predictors directly without the need for a ground truth.

%% file: tex/main.tex
\section{The \approach Algorithm}
\label{sec:main}

On a high level, \approach follows the structure of differential testing~\cite{mckeeman98}:
an \approach \emph{campaign} searches for inconsistencies between a fixed pair of throughput predictors and reports them as \emph{discoveries}.
We generate valid input basic blocks, give them to both tools under investigation, and compare their results.
The throughput predictors are required to support a common instruction set architecture (ISA), \ie, they need to have compatible input formats.
Given a basic block, they should output a real-valued estimate for the number of cycles required for its execution or report an error.

Differential testing is a natural fit to overcome the problems when comparing basic block throughput predictors described in \autoref{sec:background}.
No assumptions need to be made about the ground truth.
Differences in the assumptions of the tools under investigation are visible as inconsistencies for basic blocks that are affected by these assumptions.

Valuable insight can be gained from comparing the results of a throughput predictor to measurements on the actual hardware rather than other predictors.
\approach naturally supports this, by using a microbenchmarking tool as one of the tools under investigation.\footnote{We explore this in a case study in \autoref{ssec:cs-llvmmca}.}
However, even if microbenchmarking is very close to the actual hardware, it still makes several assumptions that may not hold when running actual code, for instance by initializing registers and memory with specific values.
Therefore, we use the perspective of differential testing for such comparisons with hardware measurements to acknowledge that the involved microbenchmarking tool is also influenced by assumptions and not a definitive ground truth.

As shown in \autoref{sec:intro}, a key challenge for \approach is that inconsistencies are so common in the random samples that just reporting all basic blocks with inconsistencies leads to an impractical number of reports.
Therefore, we center the algorithm around the idea of \emph{abstract basic blocks}, or \emph{abstract blocks} for short: compact representations characterizing sets of basic blocks by common properties.
\approach aims for several goals to make the reported abstract blocks useful:
\begin{itemize}[leftmargin=*]
  \item The reported basic blocks should be \emph{concise}, \ie, not contain instructions that are irrelevant to the underlying problem.
    This makes them easy to interpret.
  \item Each discovery should be \emph{general} by representing as many relevant basic blocks as possible.
    The more general the discoveries are, the fewer of them need to be inspected.
  \item The discoveries should be \emph{pertinent}, \ie, not represent basic blocks that do not exhibit inconsistencies in the tools under investigation.
    Significant numbers of such cases would make the characterization unreliable.
\end{itemize}
Since the results of \approach are used to hint at existing problems or show limitations of the tools -- rather than, \eg, proving the absence of inconsistencies -- none of these goals are strict formal requirements.
This fact allows us to employ approximations rather than heavy formal machinery at several points in the following sections.

\approach's high-level structure, serving as a table of contents for the remainder of this section, is shown in \autoref{algo:discovery}.
We randomly sample a basic block and check whether it is \emph{interesting}, \ie, if the throughput predictors under test exhibit an inconsistency (ll.3-5).
We minimize interesting basic blocks (l.6) by greedily removing as many instructions as we can while keeping the block interesting.
If the minimized basic block is already represented by a previously discovered abstract block, we do not need to further investigate it (ll.7-8).
Otherwise, the basic block is generalized to an abstract block, which is then noted as a new discovery (ll.9-10).

We repeat this process until some termination condition is reached (l.2), \eg, a time budget is expired or a number of subsequent samples did not produce new discoveries.
Finally, we check for each discovered abstract block~$\mathit{a}$ whether there is a subsequent one whose represented basic blocks include all of $\mathit{a}$.
Such subsumed discoveries provide only redundant information and are therefore filtered from the results (l.11).

\begin{figure}
  \centering
  \begin{minipage}{.7\linewidth}
    \begin{algorithm}[H]
      $\mathit{discoveries} \leftarrow \{\}$\;
      \While{\normalfont termination condition not reached}{
        $\mathit{candidate} \leftarrow \mathit{sampleBB}()$ \tcp*{\autoref{ssec:sampling}}
        \If(\tcp*[f]{\autoref{ssec:interestingness}}){\normalfont $\mathit{candidate}$ is not interesting } {
          \Continue\;
        }
        $\mathit{minBB} \leftarrow \mathit{minimize}(\mathit{candidate})$\;
        \If{\normalfont any $d\in \mathit{discoveries}$ subsumes $\mathit{minBB}$ } {
          \Continue \tcp*{\autoref{ssec:subsumption}}
        }
        $\mathit{newDisc} \leftarrow \mathit{generalize}(\mathit{minBB})$ \tcp*{\autoref{ssec:generalize_algo}}
        $\mathit{discoveries} \leftarrow \mathit{discovery} \cup \{\mathit{newDisc}\}$\;
      }
      \Return{$\mathit{filterSubsumed}(discoveries)$} \tcp*{\autoref{ssec:subsumption}}

      \caption{Discovering Inconsistencies.}
      \label{algo:discovery}
    \end{algorithm}

  \end{minipage}
  \Description{
    Fully described in the text.
  }
\end{figure}

In the following subsections, we describe the components of \autoref{algo:discovery} in detail as indicated in the comments.

\subsection{Interestingness Metric}
\label{ssec:interestingness}

Not every difference in the output of the tools under investigation is relevant.
Since they predict real-valued average execution times based on vastly different models, small deviations are to be expected.
Therefore, we use a more refined definition of interestingness:

A basic block is \emph{interesting} if it causes a tool under investigation to crash, or if the relative difference between their predictions $\mathit{pred}_a$ and $\mathit{pred}_b$ exceeds a specified threshold:
\[
  \frac{|\mathit{pred}_a - \mathit{pred}_b|}{\mathit{avg}(\mathit{pred}_a, \mathit{pred}_b)} = \frac{|\mathit{pred}_a - \mathit{pred}_b| \cdot 2}{\mathit{pred}_a + \mathit{pred}_b} > \mathit{threshold}
\]

As we cannot assume any of the predictions to be the \enquote{correct} one, this definition normalizes the absolute difference between the predictions by their arithmetic mean.
The interestingness threshold is a parameter of the algorithm that influences what inconsistencies are found.

Other definitions of interestingness are conceivable and may be useful.
For example, our \approach implementation also provides an alternative metric based on the absolute difference:
\[
  |\mathit{pred}_a - \mathit{pred}_b| > \mathit{threshold}
\]

Which metric is the most suitable depends on the inconsistencies we are searching for.
The relative difference is effective for focusing on interesting inconsistencies when the predicted numbers of cycles grow larger.
With the absolute difference metrics, it is easier to investigate inconsistencies of a few cycles for short-running basic blocks.

\subsection{Basic Block Abstraction}
\label{ssec:bb_abstraction}

We describe our representation of sets of basic blocks as an instance of abstract interpretation~\cite{cousot77}.
Abstract interpretation is a technique commonly used in static program analysis.
It provides a theoretical framework to over-approximate the set of possible behaviors of a program.
A key insight of the technique is to represent subsets of conceivable program behaviors (denoted as elements of the \emph{concrete domain}) by elements of an \emph{abstract domain}.
This is beneficial since the concrete domain of sets of program behaviors is generally too large to work with, whereas the abstract domain can be compact.

In \approach, we apply this notion of abstraction to a different application domain:
instead of program behaviors, we abstract basic blocks.
Therefore, our concrete domain contains sets of basic blocks and the abstract domain represents features of these basic blocks such as instruction mnemonics, use of memory, and operand dependencies.

Formally, our concrete domain~$\concdomain$ is the power set of the set~$\bbs$ of sequences of instructions in an ISA:
\[\concdomain \definedas \powset{\bbs} \text{\qquad with $\bbs \definedas \insns^+$}\]
An abstract domain~$\absdomain$ is a set with a partial order $\sqsubseteq$ that relates domain elements by their generality:
if an element~$a$ is larger than another~$b$, it represents at least all elements of the concrete domain that~$b$ represents.
While the abstract domain may be infinite, we do not allow it to have infinite sequences of strictly more general elements.

As usual in abstract interpretation, the \approach algorithm works independently of the specific abstract domain.
The abstract domain only needs to relate to the concrete domain via two functions:
a \emph{concretization function} $\gamma : \absdomain \rightarrow \concdomain$ and a \emph{representation function} $\beta : \bbs \rightarrow \absdomain$.

\begin{figure}
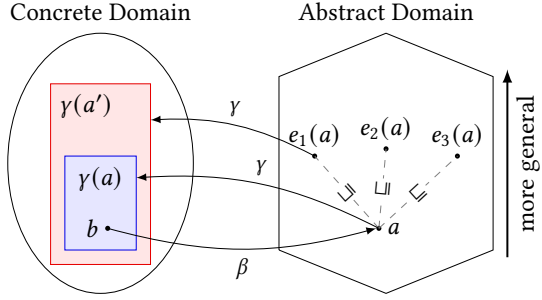

  \centering
  \scalebox{0.95}{
  \includestandalone{pics/tikz/domains}
  }
  \caption{Relationships between elements of concrete and abstract domain. }
  \label{fig:domains}
  \Description{
    The following relationships are shown:
    Starting with a concrete element b, abstracting with beta and then concretizing the result with gamma leads to a set of concrete elements that includes b.
    Starting with an abstract element a, applying an expansion and then conretizing the result leads to a set of concrete elements that fully contains the concretization of a.
    Applying an expansion to a yields a larger abstract value (with respect to the defined order).
  }
\end{figure}

The concretization function~$\gamma$ maps each element of the abstract domain to a set containing all basic blocks that it represents.
Conversely, the representation function~$\beta$ maps single basic blocks to a representation in the abstract domain.%
\footnote{We require $\beta$ instead of the more common \emph{abstraction function} $\alpha : \concdomain \rightarrow \absdomain$ since it tends to be easier to define and since our generalization algorithm only ever needs to abstract single concrete basic blocks.
  With an abstraction function~$\alpha$, $\beta$ could be defined as $\beta(x) = \alpha(\{x\})$. }
\autoref{fig:domains} visualizes these functions and the constraints imposed on them to constitute an abstract domain:
\begin{gather}
  \forall b \in \bbs.~ b \in \gamma(\beta(b))\label{eq:soundness_constraint}\\
  \forall a, a' \in \absdomain.~ a \sqsubseteq a' \Rightarrow \gamma(a) \subseteq \gamma(a')\label{eq:monotone_constraint}
\end{gather}

These constraints ensure that the functions are consistent with the orders of the domains:
Equation~\eqref{eq:soundness_constraint} ensures that a basic block $b$ is part of the concretization of its representative $\beta(b)$.
Equation~\eqref{eq:monotone_constraint} requires that $\gamma$ is monotone w.r.t.\ the domain orders, \ie, that if one abstract block is more general than another, it represents more concrete basic blocks.

While $\beta$ is often straightforward to implement, $\gamma$ is unwieldy:
if implemented explicitly, it would need to produce very large sets of concrete basic blocks.
To avoid this problem, abstract domains in \approach do not come with an explicit concretization function, but with a \emph{concretization sampler} $\widetilde\gamma(a)$ that randomly samples a basic block from $\gamma(a)$.

Our algorithm does not require an explicit generality relation $\sqsubseteq$ either.
Instead, we use a set~$\mathit{Exps} \subseteq \absdomain \rightharpoonup \absdomain$ of partial \emph{expansion} functions that each map abstract blocks to their immediate successors in the generality relation.

In practice, these expansion functions each describe a way to modify abstract blocks in order to obtain a slightly more general abstract block.
Formally, each expansion function $E\in\mathit{Exps}$ needs to be strictly ascending and monotonic:
\begin{gather*}
  \forall a\in\fundom{E}.~ a \sqsubseteq E(a) \land a \neq E(a) \\
  \forall a, a'\in\fundom{E}.~ a \sqsubseteq a'\Rightarrow E(a) \sqsubseteq E(a')
\end{gather*}
If there is an immediate successor~$a'$ to $a$ in the generality order, there should be an expansion function $E \in \mathit{Exps}$ such that $E(a) = a'$.
However, we require that for each abstract block~$a$, the number of expansion functions $E \in \mathit{Exps}$ such that $a\in \fundom{E}$ is finite.

\begin{exmp}
  \label{ex:informal_ab}
For an informal notion of what an abstract block for the x86-64 instruction set architecture looks like, consider the following example:
\begin{equation}
    \fbox{
\begin{minipage}{0.60\columnwidth}
  \noindent
  Instructions:
  \begin{enumerate}[leftmargin=0.7cm, topsep=0.0cm]
    \item mnemonic: \texttt{mov}; mem(ory usage): R (= Read)
    \item mnemonic: \texttt{add}; category: arithmetic; mem: R+W
  \end{enumerate}

  \noindent
  Aliasing:
  \begin{itemize}[leftmargin=0.6cm, topsep=0.0cm]
    \item operand 1 of insn 1 must alias with operand 2 of insn 2
  \end{itemize}
\end{minipage}
} \tag{AB1}\label{ex:ab1}
\end{equation}
  This abstract block represents all basic blocks consisting of two instructions that satisfy constraints on their mnemonics, their category, their use of memory, and the aliasing of their operands.\footnote{We use the term \enquote{alias} here for instruction operands that refer to the same data. It is therefore not restricted to memory operands, but also refers to (fully or partially) overlapping register operands.}
The following is one of the concrete basic blocks represented by the above abstract block:
\begin{center}
  \begin{minipage}{0.4\columnwidth}
  \begin{lstlisting}
mov rbx, [rdx + 42]
add [r8], rbx
  \end{lstlisting}
  \end{minipage}
\end{center}
The mnemonics fit their constraints, the first instruction only reads memory (at the location \texttt{rdx} + 42), whereas the second one reads and writes memory at \texttt{r8}.
The aliasing constraint is satisfied by using the common \texttt{rbx} register.

An expansion function could for example drop the constraint on the mnemonic of the second instruction.
The result of this expansion function is the following abstract block:
\begin{center}
\fbox{
\begin{minipage}{0.60\columnwidth}
  \noindent
  Instructions:
  \begin{enumerate}[leftmargin=0.7cm, topsep=0.0cm]
    \item mnemonic: \texttt{mov}; mem: R
    \item category: arithmetic; mem: R+W
  \end{enumerate}

  \noindent
  Aliasing:
  \begin{itemize}[leftmargin=0.6cm, topsep=0.0cm]
    \item operand 1 of insn 1 must alias with operand 2 of insn 2
  \end{itemize}
\end{minipage}
}
\end{center}
It represents all basic blocks represented by the previous one, but is less specific:
all arithmetic instructions are now allowed as the second instruction.
\end{exmp}

Next, we describe how \approach automatically generalizes interesting basic blocks to concise and pertinent abstract blocks.
With this generalization algorithm in mind, we then formalize the details of a modular abstract domain for the x86-64 instruction set architecture in \autoref{ssec:absdom}.

\subsection{Generalization Algorithm}
\label{ssec:generalize_algo}

\approach generalizes an interesting basic block~$b$ as shown in \autoref{algo:generalization}.
The first result candidate is the representative~$\beta(b)$, an abstract block that represents the given basic block~$b$ as specifically as possible in the abstract domain (l.1).
After validating that the initial candidate is interesting (l.2), we choose an expansion to make the candidate more general (ll.7-8).
If the expanded abstract block is still interesting, we use it as a new candidate (l.9).
Otherwise, we note this expansion as rejected (l.10) and choose a different one.
As expansions are monotonic and ascending, a once rejected expansion cannot be useful later in generalization.

Once no expansion is left, we return the now general and still pertinent candidate (ll.5-6).
Termination is guaranteed as the abstract domain has no infinite ascending chains and the set of expansions that apply to an abstract block is finite.

\begin{figure}
  \centering
  \begin{minipage}{.62\linewidth}
    \begin{algorithm}[H]
      \KwIn{basic block $b$}
      $\mathit{absBB} \leftarrow \beta(b)$\;
      \lIf{\normalfont$\mathit{absBB}$ is not interesting}{\Return $b$}\label{algo:line:earlycheck}
      $\mathit{rejected} \leftarrow \{\}$\;

      \While{True}{
        $\mathit{avail} \leftarrow \{ E \in \mathit{Exps} \mid \mathit{absBB} \in \fundom{E} \} \setminus \mathit{rejected}$\;

        \lIf{$\mathit{avail} = \{\}$}{
          \Return $absBB$
        }
        $\mathit{exp} \leftarrow \mathit{choose}(\mathit{avail})$\;\label{algo:line:choose_exp}

        $\mathit{t} \leftarrow \mathit{exp}(\mathit{absBB})$\;

        \lIf{\normalfont $\mathit{t}$ is interesting}{
          $\mathit{absBB} \leftarrow \mathit{t}$
        }
        \lElse{
          $\mathit{rejected} \leftarrow \mathit{rejected} \cup \{\mathit{exp}\}$
        }
      }

      \caption{Generalization Algorithm.}
      \label{algo:generalization}
    \end{algorithm}
  \end{minipage}
  \Description{
    Fully described in the text.
  }
\end{figure}

We extend our definition of interestingness (\autoref{ssec:interestingness}) from basic blocks to abstract blocks for this algorithm.
Ideally, an abstract block should be deemed interesting if all represented basic blocks are interesting.
As this is prohibitively expensive to check, we approximate this property:
we randomly sample represented basic blocks and consider the abstract block interesting if all samples are interesting.
The number of samples is a parameter of \approach.

\begin{exmp}
  \label{ex:generalization}

Assume that the predictors under test disagree on the latency of reading a value from memory that was written immediately before.
\autoref{fig:example_gen} visualizes a run of the generalization algorithm for this problem.

The first hypothesis for an abstract block (1) is the representation of a concrete basic block exhibiting this behavior.
In the next step, the algorithm expands the aliasing requirement and reaches abstract block (2).
With $\top$, we denote that the component is unconstrained.
Since the sampled basic blocks are then no longer restricted to using the same memory location, they are not uniformly interesting, which causes this expansion to be rejected.
When we expand the mnemonic of the second instruction (3), the abstract block continues to only cover interesting basic blocks.
Allowing any of the instructions to not use memory (4,5) leads to more rejections.
Finally, this leaves only the mnemonic of the first instruction to be expanded (6).
After that, all components of the abstract blocks are either $\top$ or only affected by rejected expansions.
Hence, the algorithm terminates returning abstract block (6).

\begin{figure}
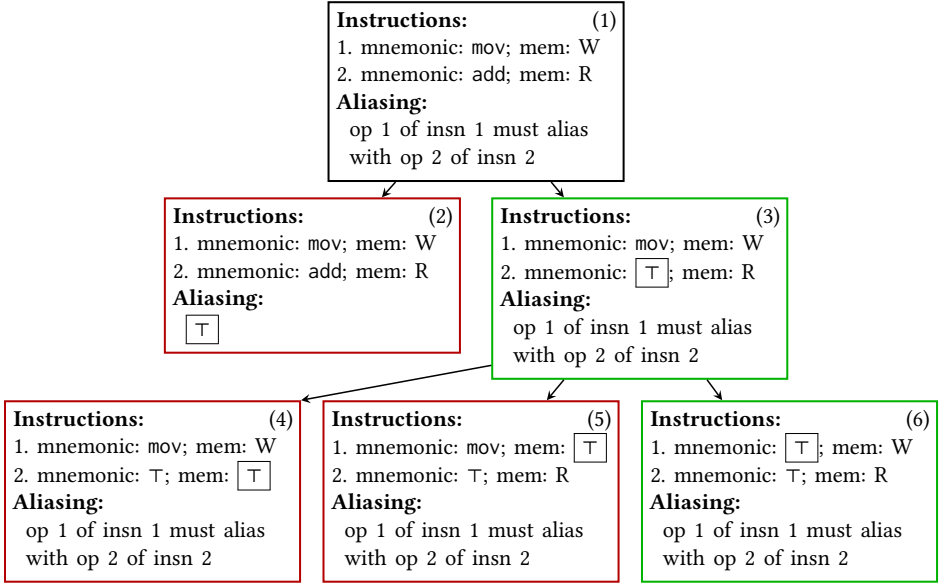

  \centering
  \scalebox{0.95}{
  \includestandalone{pics/tikz/gen_tree}
  }
  \caption{An example generalization tree. }
  \label{fig:example_gen}
  \Description{
    Fully described in the text.
  }
\end{figure}
\end{exmp}

The order in which expansions are chosen in the generalization algorithm (line~\ref{algo:line:choose_exp} in \autoref{algo:generalization}) affects the result.
We approximate an optimal expansion order by generalizing each candidate several times with different random expansion orders.
Since we prune subsumed discoveries from the results, we can try arbitrarily many different expansion orders without degrading the quality of our discovery results.

The straightforward nature of the generalization algorithm is helpful when interpreting \approach's results.
As \citet{parnin2011} noted, tools that automatically find bugs and present them only abstractly to users do not necessarily help fix the bugs.
They formulate the observation that \enquote{[p]roviding overviews that cluster results and explanations that include data values [and] test case information [\dots] could make faults easier to identify and tools ultimately more effective.} \citep[Section 6.1]{parnin2011}

Our generalization algorithm naturally produces such clustered results and explanations in the form of a generalization decision tree like the one in \autoref{fig:example_gen}.
Each decision in this tree is justified by the set of basic blocks that was sampled and evaluated to gauge their interestingness.
Our implementation of \approach therefore includes a graphical interface to inspect the generalization trees of its discoveries to provide users with detailed information and concrete debuggable inputs.

Particularly insightful are the basic blocks that justify the rejection of an expansion.
They highlight the limits of an inconsistency's scope in a way that a plain clustering of inconsistent basic blocks could not.
We will demonstrate in \autoref{sec:eval} how we can use such results to identify behaviors of throughput predictors that run counter to common expectations.

\subsection{Our Abstract Domain}
\label{ssec:absdom}

We now define an abstract domain for the x86-64 instruction set architecture as this ISA is supported by most available basic block throughput predictors.
The \approach algorithm is not conceptually limited to this ISA and similar domains can be designed for, \eg, ARM architectures.

\subsubsection*{Top-Level Abstraction}
Our abstract domain~$(\absdomain, \sqsubseteq_\absdomain)$ separates constraints on the individual instructions of the represented basic blocks from constraints on how they interact via their operands.
An abstract block thus consists of a sequence of abstract instructions and an abstract alias information:
\[
  \absdomain \definedas \absdomain_\idxinsn^+ \times \absdomain_\idxalias
\]
The partial order $\sqsubseteq_\absdomain$ only relates abstract blocks with the same number of abstract instructions and is defined through partial orders~$\sqsubseteq_\idxinsn$ and $\sqsubseteq_\idxalias$ among its components:
\begin{gather*}
  \begin{aligned}
    (x_\idxinsn, x_\idxalias) \sqsubseteq_\absdomain  (y_\idxinsn, y_\idxalias) \definedeq~& x_\idxalias \sqsubseteq_\idxalias y_\idxalias \land \abs{x_\idxinsn} = \abs{y_\idxinsn}\\
                &\land (\forall 1 \leq k \leq \abs{x_\idxinsn} .~x_\idxinsn[k] \sqsubseteq_\idxinsn y_\idxinsn[k])
  \end{aligned}
\end{gather*}
In a similar way, the concretization~$\gamma_\absdomain$ and representation~$\beta_\absdomain$ functions are defined based on per-component-functions:
\begin{align*}
  b \in \gamma_\absdomain( (a_{\idxinsn}, a_\idxalias)) &\definedeq  b \in \gamma_\idxalias(a_\idxalias) \land |a_\idxinsn| = |b|
    \land (\forall 1 \leq k \leq |a_\idxinsn| .~ b[k] \in \gamma_\idxinsn(a_\idxinsn[k])) \\
  \beta_\absdomain(b) &\definedas ([\beta_\idxinsn(i) \mid i \in b], \beta_\idxalias(b))
\end{align*}

Expansion functions expand one component of the abstract block, \ie, an abstract instruction or the aliasing abstraction, and leave all other components untouched:
\begin{align*}
  \mathit{Exps}_\absdomain \definedas&
    \set{\lambda (a_\idxinsn, a_\idxalias).\ (a_\idxinsn[k \mapsto E(a_\idxinsn[k])], a_\idxalias) \text{ if $k \leq \abs{a_\idxinsn} \land a_\idxinsn[k]\in\fundom{E}$}
    \given  k\in \nat, E\in\mathit{Exps}_\idxinsn}\\
     &\mathrel{\cup} \set{ \lambda (a_\idxinsn, a_\idxalias).\ (a_\idxinsn, E(a_\idxalias)) \text{ if $a_\idxalias\in\fundom{E}$} \given E\in\mathit{Exps}_\idxalias }
\end{align*}

\subsubsection*{Instruction Abstraction}
The set~$\absdomain_\idxinsn$ contains abstract instructions that describe sets of \emph{instruction schemes}.
An instruction scheme (or instruction variant) is an instruction representation that is parametric in its operands:
it specifies the width and kind of the operands, but does not specify an actual register, memory operand, or immediate value.
For example, the instruction scheme
\begin{center}
\texttt{add \textlangle GPR:64\textrangle, \textlangle MEM:64\textrangle}
\end{center}
describes 64-bit addition instructions with a register as the first and a memory reference as the second operand.

We extract the instruction schemes for the x86-64 instruction set architecture from uops.info~\cite{abel19uopsinfo}.
Additionally, we collect for each instruction scheme several features such as the mnemonic, the operand types, whether and how it uses memory, and to which instruction category and ISA extension it belongs.
Our domain groups instructions through constraints on these features.
The form of these constraints for a feature~$f$ is determined by its feature abstraction~$\absdomain_F[f]$.
\autoref{fig:feature_domains} introduces the feature abstractions that we use for our abstract domain.

For simple features like the category and ISA extension to which the instruction belongs or the presence of a \texttt{lock} or \texttt{rep} prefix, we use the \emph{singleton} abstraction.
It expresses that all represented instruction schemes have a specific value for the feature.

For the mnemonic, we use the \emph{edit distances} abstraction.
It constrains the represented mnemonics by an upper bound $\mathit{d}$ on the Levenshtein distance from a base string~$B$.
If they share their base, abstract values are ordered by their value of~$\mathit{d}$, which is limited by a maximum bound (here: 3).
This abstraction allows \approach to group instructions with similar mnemonics:
an abstract value representing only \texttt{vaddpd} (an addition for vectors of double-precision floats) can be expanded to also represent \texttt{vaddps} (the same operation with single-precision) and the scalar versions \texttt{vaddsd} and \texttt{vaddss}.
The edit distance abstraction is heuristic in nature: neither do all similar instructions have similar mnemonics nor are all instructions with similar mnemonics similar themselves.
We nevertheless found this abstraction to be helpful in practice since, for instance, mnemonic suffixes that rarely affect the instruction's performance behavior  are common in modern ISAs (\eg, specifying floating-point format and vector width, or the condition for conditional move instructions).

We use the \emph{log sizes} abstraction for the (multi-)set of micro operations required to execute an instruction.
\approach can therefore group instructions by their complexity:
an abstract value represents all instruction schemes that are decomposed into less than $2^k$ \uops for a certain $k$.
To avoid infinite ascending chains of abstractions, a maximal value for $k$ is a parameter of this abstraction (here: 5).

Whether and how an instruction accesses memory affects its performance significantly.
Our domain therefore uses the fine-grained \emph{subset-or-none} abstraction with subsets of $\{\mathsf{R}, \mathsf{W}, \mathsf{Size}:n\}$ to represent memory usage.
This enables \approach to relax constraints on memory usage step by step.
An abstract instruction representing only instructions that \textsf{R}ead and \textsf{W}rite $n$ bits in memory can be expanded by dropping any of these constraints.
The expanded abstract instruction might represent all instructions that at least read $n$ bits from memory.
With $\mathsf{DefNone}$, only instructions that do not access memory are represented.
We also use this abstraction for the set of operand types that occur in the instruction schemes.

\begin{table*}
  \caption{Feature domains used in \approach.
    The domains are shown as Hasse Diagrams, where the partial order is indicated through the lines:
    if $x$ is connected to $y$ and $y$ is closer to the top, $x \sqsubseteq_F y$ holds.}
  \label{fig:feature_domains}
  \small

  \begin{tabular}{ccccc}
    \toprule
    Domain & Hasse Diagram & Used for & {$\gamma^{(f)}_{X}(\mathit{av})$ for $\mathit{av} \neq \top$} & $\beta^{(f)}_{X}(i)$ \vphantom{\Large X\Huge g} \\
    \toprule
    Singletons & %
      \parbox[c]{1.5cm}{\centering \makebox[0pt]{ \scalebox{0.95}{ \includestandalone{pics/tikz/abstraction_singleton}} }\par} & %
      \parbox{1.4cm}{\centering Category,\\ISA-Set,\\Prefixes} & %
      $\{i \mid f(i) = \mathit{av}\}$ & %
      $f(i)$ \\
    \midrule
    \parbox{1.4cm}{\centering Edit\\Distances} & %
      \parbox[c]{3cm}{\centering \makebox[0pt]{ \scalebox{0.95}{ \includestandalone{pics/tikz/abstraction_edits}} }\par} & %
      Mnemonic & %
      $\{i \mid \mathit{dist}(f(i), \mathit{B}) \leq \mathit{d}\}$ & %
      $(\mathit{B}: f(i), \mathit{d}: 0)$ \\
    \midrule
    Log Sizes & %
      \parbox[c]{1cm}{\centering \makebox[0pt]{ \scalebox{0.95}{ \includestandalone{pics/tikz/abstraction_logsizes}} }\par} & %
      \parbox[c]{1.4cm}{\centering Number\\of \uops} & %
      $\{i \mid \abs{f(i)} < 2^\mathit{av}\}$& %
      $\floor*{\log_2(\abs{f(i)} + 1)}$ \\
    \midrule
    \parbox[c]{1.4cm}{\centering Subset-\\or-None} & %
      \parbox[c]{4.2cm}{\centering \makebox[0pt]{ \scalebox{0.95}{ \includestandalone{pics/tikz/abstraction_subset_or_dn}} }\par} & %
      \parbox[c]{1.4cm}{\centering Memory Usage,\\Operand Types} & %
      \parbox{2.4cm}{
        if $\mathit{av} = \mathsf{DefNone}$:\\
        \hspace*{1em}$\{i \mid f(i) = \emptyset\}$ \\
        otherwise: \\
        \hspace*{1em}$\{i \mid f(i) \supseteq \mathit{av}\}$
      } & %
      \parbox{2cm}{
        if $f(i) = \emptyset$:\\
        \hspace*{1em}$\mathsf{DefNone}$\\
        otherwise:\\
        \hspace*{1em}$f(i)$
      }\\
    \midrule
  \end{tabular}
\end{table*}

Formally, abstract instructions are tuples of feature abstraction~$\absdomain_F[f_i]$ elements for each considered feature~$f_i$:
\[
  \absdomain_\idxinsn \definedas \absdomain_{\mathit{F}}[\mathit{f}_1] \times\cdots\times \absdomain_{\mathit{F}}[\mathit{f}_N]
\]

The partial order among abstract instructions relies on the partial orders of the involved feature abstractions:
\[
  (x_1, \cdots, x_N) \sqsubseteq_\idxinsn (y_1, \cdots, y_N) \definedeq \bigland_{i\in[1, N]} x_i \sqsubseteq y_i
\]

The last two columns of \autoref{fig:feature_domains} define the feature concretization and representation functions~$\gamma^{(f)}_X$ and~$\beta^{(f)}_X$ for each feature abstraction~$X$.
They are parameterized by the feature~$f$ for which they are used and refer with $f(i)$ to the value of the instruction~$i$ for this feature.

The feature concretization functions~$\gamma^{(f)}_X$ map \emph{abstract values} from the feature abstraction~$X$ to their set of represented instructions~$i$.
All feature abstractions have a maximal abstract value $\top$, which represents the absence of any constraint.
Therefore its concretization is the same for every domain: the entire set of available instruction schemes.
The feature representation functions $\beta^{(f)}_X$ define the value from the feature abstraction that best describes an instruction with the value~$f(i)$ for the feature~$f$.

We use these feature concretization and representation functions to define the instruction-wise functions:
an abstract instruction imposes the conjunction of the per-feature constraints on the represented instructions.
Therefore, it concretizes to the intersection of the per-feature concretizations~$\gamma_{\absdomain_{F}[f]}^{(f)}$ applied to the abstract instruction's component~$\mathit{ai}[f]$ for each feature~$f$:
\[
  \gamma_\idxinsn(\mathit{ai}) \definedas \bigcap_{f\in \mathit{Features}} \gamma_{\absdomain_{F}[f]}^{(f)}(\mathit{ai}[f])
\]
To obtain a representative abstract instruction for a concrete one, we apply the representation functions for each feature:
\[
  \beta_\idxinsn(i) \definedas \left(\beta^{(f_1)}_{\absdomain_{F}[f_1]}(i), \dots, \beta^{(f_N)}_{\absdomain_{F}[f_N]}(i)\right)
\]
Analogously, an expansion function for an abstract instruction~$i$ takes a non-$\top$ component and replaces it with one of its immediate successors in the generalization order:
\begin{gather*}
  \mathit{Exps}_\idxinsn \definedas \set{ \lambda \mathit{ai}.~ \mathit{ai}[f \mapsto y] \text{ if $\mathit{ai}[f] = x$ } \given
    f\in\mathit{Features}, x\in\absdomain_{F}[f]\setminus \{\top\}, y \text{ succeeds $x$ in $\sqsubseteq_{\absdomain_{F}[f]}$} }
\end{gather*}

\subsubsection*{Aliasing Abstraction}
The subcomponent~$\absdomain_\idxalias$ represents aliasing constraints among operands of instructions.
We refer to an operand of an instruction via a pair~$(\mathit{idx}_i,\mathit{idx}_o) \in \mathit{Idx} \definedas (\mathit{InsnIdx} \mathop{\times} \mathit{OpIdx})$ of indexes into the sequence of instructions and into the sequence of operands of the instruction.
The operand $\mathit{idx}_o$ of instruction $\mathit{idx}_i$ in the basic block $b$ is denoted as $b[(\mathit{idx}_i, \mathit{idx}_o)]$.

An aliasing constraint for a pair of such instruction operand designators can state that they \emph{must} or \emph{must not} alias, or that no constraint applies (denoted as~$\top$).
The aliasing subcomponent is therefore defined as a mapping as follows:
\begin{gather*}
  \absdomain_\idxalias \definedas (\mathit{Idx} \times \mathit{Idx}) \rightarrow \{\texttt{must}, \texttt{mustnot}, \top\}
\end{gather*}
An aliasing information~$g$ is more general than another~$h$ if $g$ imposes the same or a weaker constraint than $h$ for \emph{every} pair of operands:
\begin{gather*}
  \begin{multlined}
    h \sqsubseteq_\idxalias g \definedeq \forall x \in (\mathit{Idx} \times \mathit{Idx}).\ g(x) = \top \lor h(x) = g(x)
  \end{multlined}
\end{gather*}

This component of the abstraction is more intricate than one might expect at first:
the instruction abstraction can represent sets of vastly different instruction sequences.
One abstract instruction could for example represent a 2-operand integer addition operation and a 3-operand floating point addition.
Consequently, the aliasing abstraction needs to handle cases where operands do not match, \ie, cannot possibly alias, or where they are not present at all in some of the represented basic blocks.

We handle these cases with a concretization function that only applies constraints on operands that match and are present in the concrete basic block:
\begin{equation*}
\begin{multlined}
  b \in \gamma_\idxalias(h) \definedeq
  \smashoperator{\bigland_{((\mathit{i}_1, \mathit{i}_2) \mapsto x) \in \mathit{h}}} \Big(\mathit{existAndMatch}\big(b[\mathit{i}_1], b[\mathit{i}_2]\big) \\
  \Rightarrow \Big(x = \top \lor \left(x = \texttt{must} \land \mathit{doAlias}\big(b[i_1], b[i_2]\big) \right)
  \lor \left(x = \texttt{mustnot} \land \lnot \mathit{doAlias}\big(b[i_1], b[i_2]\big)\right)\Big)\Big)
\end{multlined}
\end{equation*}

The representation function~$\beta_\idxalias$ is defined to capture the must-alias and must-not-alias relations between the matching operands of the concrete basic block:
\begin{gather*}
  \begin{multlined}
  \beta_\idxalias(b) \definedas \lambda (i_1, i_2).
    \begin{cases}
      \texttt{must} & \parbox{17em}{if $b[i_1], b[i_2]$ exist, match, and alias}\\
      \texttt{mustnot} & \parbox{17em}{if $b[i_1], b[i_2]$ exist, match, and do not alias}\\
      \top & \text{otherwise}
    \end{cases}
  \end{multlined}
\end{gather*}

Deciding for a pair of register operands whether they alias is straightforward:
they alias if and only if they are the same or if one is a sub-register of the other.\footnote{We do not consider the legacy floating point extensions x87 and MMX; register aliasing is more complicated for the x87 register stack.}
Whether memory operands alias depends on the values of registers.
We approximate this by considering two memory operands aliasing if they are identical, and not aliasing otherwise.
In general, this is not a sound approximation: two memory operands can look entirely different but refer to the same address or vice versa.
It is only sound for our use case because the basic block sampling method described in the following section manages memory operands such that they alias if and only if they are syntactically identical.

The expansion functions for the aliasing component of an abstract block each replace a non-$\top$ entry in the aliasing abstraction with $\top$:
\[
  \mathit{Exps}_\idxalias \definedas \set{ \lambda h.~ h[x\mapsto \top] \text{ if $h(x) \neq \top$}  \given x \in \mathit{Idx} }
\]

\subsection{Sampling Represented Basic Blocks}
\label{ssec:sampling}
Our generalization algorithm relies on a method~$\widetilde\gamma$ to randomly sample basic blocks that are represented by a given abstract block.
For arbitrary elements of our abstract domain, this is a hard problem:
sampling a block that fulfills the aliasing constraints essentially corresponds to a graph-coloring register allocation problem~\cite{graphcol} because concrete registers have to be found that comply with the aliasing constraints.
Since there are no restrictions on these constraints, arbitrary interference graphs can emerge in general which renders sampling NP-hard in theory.

As the concretization sampler~$\widetilde\gamma$ is a very common operation in \approach's generalization algorithm, we do not implement a complete solution to the NP-hard sampling problem.
Instead, we proceed greedily as follows:
\begin{enumerate}[leftmargin=*]
  \item For each abstract instruction, choose a represented instruction scheme.
  \item If the schemes have fixed operands\footnote{For example, shifts in x86 use the \texttt{c} register for their shift amount.}, select those and set all related must-alias operands accordingly.
  \item Repeatedly: Where an operand is not yet selected, choose one that is not forbidden through must-not-alias constraints. Set all must-alias operands accordingly.
\end{enumerate}
We restrict what registers may be used as register operands to have distinct registers available for the base registers of memory operands that cannot be overwritten.
If two memory operands are required to alias, we instantiate them with the same combination of base register and displacement.
In case of a no-alias constraint on memory operands, we use different combinations of base register and displacement.

If, at any point in this algorithm, no selection is possible without violating the alias constraints or the requirements of the instruction schemes, the sampling fails and needs to be repeated.
For an example, consider the following abstract block for the x86 ISA with two unconstrained abstract instructions and a must-not alias constraint:
\begin{center}
    \fbox{
\begin{minipage}{0.70\columnwidth}
  \noindent
  Instructions:
  \begin{enumerate}[leftmargin=0.7cm, topsep=0.0cm]
    \item $\top$
    \item $\top$
  \end{enumerate}

  \noindent
  Aliasing:
  \begin{itemize}[leftmargin=0.6cm, topsep=0.0cm]
    \item operand 2 of insn 1 must not alias with operand 2 of insn 2
  \end{itemize}
\end{minipage}
}
\end{center}
If, in the first step of the sampling algorithm, we choose shift instructions with variable shift amount for both instructions, sampling will fail:
then both instructions need to have register \texttt{c} as second operand (for the shift operand), which would violate the alias constraint.

In practice, sampling rarely fails for the short instruction sequences that we sample: it affected 0.01\% of the ca.\ $4.8\times10^6$ sampling operations in the campaigns presented in \autoref{ssec:effectiveness}.%
\footnote{An alternative approach would be to encode the aliasing constraints as SMT formulae and use, \eg, the approach of \citet{dutra19smtsampling} to sample satisfying solutions.
In comparison to our approach this would eliminate the chance of sampling errors at the cost of an increased execution time of the sampling steps.}

\subsection{Checking for Subsumption}
\label{ssec:subsumption}

In \autoref{algo:discovery}, we check whether concrete or abstract blocks are subsumed by an abstract block to avoid unnecessary generalizations and to prune irrelevant discoveries.
The fixed number and positions of instructions in our abstract domain ease sampling basic blocks, but they hinder us here:
the concretization~$\gamma_\absdomain(a)$ of an abstract block $a$ does not contain basic blocks that we would like to consider subsumed by $a$.
The following basic block would not be included in the concretization of abstract block~\ref{ex:ab1} from Example~\ref{ex:informal_ab}:
\begin{center}
  \begin{minipage}{0.45\columnwidth}
  \begin{lstlisting}
add [r8], rbx; mov rbx, [rdx + 42]
  \end{lstlisting}
  \end{minipage}
\end{center}
However, the instructions here are the same as in the example represented by~\ref{ex:ab1}, only in a different order that has no impact on the block's sustained throughput:
the throughput is determined by the rate at which the basic block can be executed repeatedly for an indefinite number of iterations.
What determines the throughput of a basic block $\mathit{bb}$ is therefore the trace of instructions resulting from repeating $\mathit{bb}$ a large number of times.
When a basic block $\mathit{bb}'$ results from rotating $\mathit{bb}$ (\ie, removing a sequence of instructions from the beginning and appending it to the end), its trace differs from the one of $\mathit{bb}$ only by short pre- and suffixes whose influence on the execution time vanishes with a growing number of repetitions.

Similarly, if this basic block exhibits an inconsistency in the predictions, it is likely to have the same reason as~\ref{ex:ab1}:
\begin{center}
  \begin{minipage}{0.5\columnwidth}
  \begin{lstlisting}
mov rbx, [rdx + 42]; nop; add [r8], rbx
  \end{lstlisting}
  \end{minipage}
\end{center}
Yet, it is not included in ~$\gamma_\absdomain(\mathrm{AB1})$ since it contains three instead of two instructions.

We therefore do not rely on the partial order of the abstract domain to implement the subsumption checks in \autoref{algo:discovery}.
Instead, we check for the following definition:

\begin{defn}
  An abstract block $(a^1_\idxinsn, a^1_\idxalias)$ \emph{subsumes} another $(a^2_\idxinsn, a^2_\idxalias)$
  if there is a mapping~$m: I^1 \rightarrow I^2$ from the indexes~$I^1$ of the abstract instructions of $a^1_\idxinsn$ to the indexes~$I^2$ of $a^2_\idxinsn$ s.t.
  \begin{align}
    &\forall i, j \in I^1.~ m(i) \neq m(j)\tag{C1}\label{constraint_injective}\\
    &\forall i \in I^1.~ \gamma_\idxinsn(m(i)) \subseteq \gamma_\idxinsn(i)\tag{C2}\label{constraint_subset}\\
    &\begin{multlined}\forall ((i, op_1), (j, op_2) \mapsto x) \in a^1_\idxalias.
          \ x\neq\top \Rightarrow a^2_\idxalias((m(i), op_1), (m(j), op_2)) = x\end{multlined}\tag{C3}\label{constraint_alias}\\
    &\begin{multlined}\forall i \in I^1.~ \forall k \in I^2 \text{ between $m(i)$ and }
      \text{$m((i+1) \bmod \abs{I^1})$}.\ \not\exists i'.~m(i') = k\end{multlined}\tag{C4}\label{constraint_order}
  \end{align}
  An abstract block $a$ \emph{subsumes} a concrete basic block $b$ if it subsumes $\beta_\absdomain(b)$.

\end{defn}

In other words, $m$ needs to be injective (\ref{constraint_injective}) and map abstract instructions to at least as specific ones (\ref{constraint_subset}).
Furthermore, the aliasing constraints imposed by $a^2_\idxalias$ on the mapped instructions need to be at least as strong as those imposed by $a^1_\idxalias$ (\ref{constraint_alias}).
Lastly, the order of the mapped instructions $m(i)$ in $a^2_\idxinsn$ needs to be a rotation of the order of the instructions $i$ in $a^1_\idxinsn$ (\ref{constraint_order}).
All instructions of $a^1_\idxinsn$ need to have a counterpart in $a^2_\idxinsn$, but not vice versa.

We encode these constraints in a boolean formula that is satisfiable if and only if such a mapping exists and use a SAT solver to discharge them.
In an \approach campaign, subsumption checks are not numerous and in our experience, SAT solvers can solve the formulae quickly.

\subsection{Ranking Abstract Basic Blocks}
\label{ssec:discoveryranking}

When evaluating the usefulness of \approach discoveries, as well as for guiding developers interested in improving throughput predictors, it is helpful to rank abstract basic blocks by a notion of importance.
In the following, we describe three approaches to ranking abstract basic blocks that we found useful when evaluating \approach and carrying out the case studies presented in \autoref{sec:case-studies}.

\subsubsection{Ranking by Interestingness}
\label{ssec:discoveryranking-interestingness}
Every abstract block that results from \approach's generalization has been checked for interestingness.
This means that we sampled a number of represented concrete basic blocks and computed the (relative or absolute) difference between the predictions of the tools under investiation for the basic blocks for each discovery.
A natural metric of relevance of the abstract block is therefore the mean prediction difference over the set of sampled basic blocks.
The higher it is, the more dramatic is the inconsistency characterized by the abstract block.
For inputs that crash a throughput predictor, we set this metric to infinity to indicate maximal interestingness.

\subsubsection{Ranking by Generality.}
\label{ssec:discoveryranking-generality}
Inconsistencies do not need to come with large deviations in the predictions to indicate a significant difference in the tools under investigation.
We therefore use \emph{generality} as an alternative metric for ranking abstract blocks.
The idea is that we want to find discoveries that affect large classes of concrete basic blocks.

There are several conceivable options to define such a metric, which may differ in the effort required to compute them (\eg, one might sample a large number of basic blocks and check how many of them are subsumed by each discovery) and in how basic blocks are weighted (\eg, should instruction schemes with wide immediate constants be considered more general, because each possible immediate value counts as a different instruction?).

We chose a notion of generality that is inexpensive to compute and operates, like our generalization algorithm, on the granularity of instruction schemes:
an abstract block's generality is the minimal number of instruction schemes represented by any of its abstract instructions.
While this is a simplification of reality -- it ignores aliasing constraints and the number of abstract instructions in the abstract block -- this metric was instrumental to find several examples for our case studies.

\subsubsection{Maximizing the Number of Subsumed Basic Blocks.}
\label{ssec:discoveryranking-subset}

If users of \approach have a concrete set of basic blocks that they consider particularly relevant, \eg, extracted from an important benchmark set, this can be leveraged to a custom-tailored notion of generality.
For one, we can rank \approach's abstract blocks by the number of basic blocks from the set that they subsume.

An extension to this strategy is to solve the following integer linear program (ILP) to obtain a set of~$k$ discoveries from $\mathit{AbsBlocks}$ that subsume a maximally large portion of the basic block set~$B$:
\begin{align*}
  \text{maximize} \quad & \sum_{j\in B} \mathsf{BB.covered}[j] \\
  \text{subject to} \quad &\sum_{i\in \mathit{AbsBlocks}}  \hspace*{-0.2cm}\mathsf{AB.used}[i] \leq k \\
  &  \sum_{i\in \mathit{AbsBlocks}\,\land\,i\,\mathit{subsumes}\,j} \hspace*{-0.9cm} \mathsf{AB.used}[i] \geq \mathsf{BB.covered}[j] && \quad\quad\text{for all $j\in B$}\\
  &  \mathsf{AB.used}[i] \in \{0, 1\}       && \quad\quad\text{for all $i\in \mathit{AbsBlocks}$}\\
  &  \mathsf{BB.covered}[j] \in \{0, 1\}       && \quad\quad\text{for all $j\in B$}
\end{align*}

The ILP uses two groups of binary variables: an $\mathsf{AB.used}[i]$ variable for each abstract block~$i$ and a $\mathsf{BB.covered}[j]$ variable for each concrete basic block~$j$.
If one of the $\mathsf{AB.used}$ variables is~1, the corresponding abstract block is chosen as one of the $k$ maximally subsuming discoveries.
The first constraint of the ILP ensures that no more than $k$ abstract blocks are selected.
If one of the $\mathsf{BB.covered}$ variables is~1, the corresponding concrete basic block is subsumed by at least one of the chosen discoveries.
We encode this relationship with the second constraint of the ILP: $\mathsf{BB.covered}[j]$ cannot be greater than 0 unless an abstract block~$i$ that subsumes it is chosen with $\mathsf{AB.used}[i]$.
With the ILP's objective term, we require that an optimal solution covers as many concrete basic blocks as possible.

A set of abstract blocks extracted from the values of the $\mathsf{AB.used}$ variables in a solution to the ILP is a maximally diverse selection of \approach discoveries.
With an appropriate selection of the parameter~$k$, \approach's results can thus be summarized as concisely as desired.

%% file: pics/tikz/domains.tex
    \begin{tikzpicture}
        \draw (-2.5,2.1) node[align=center]{Concrete Domain};
        \draw (-2.5,0) ellipse (1.3 and 1.8);

        \draw (1.5,2.1) node[align=center]{Abstract Domain};
        \draw (0.0, -1.2) -- (1.5, -1.8) -- (3.0, -1.2) -- (3.0, 1.2) -- (1.5, 1.8) -- (0.0, 1.2) -- (0.0, -1.2);

        \draw[color=red!90!black,fill=red!10!white] (-3.2,-1.4) rectangle (-1.8,1.1);
        \coordinate[label=$\gamma(a')$] (Bp) at (-2.7,0.5);

        \draw[color=blue!90!black,fill=blue!10!white] (-3.0,-1.2) rectangle (-2.0,0.1);
        \coordinate[label=$\gamma(a)$] (B) at (-2.5,-0.5);

        \coordinate[label=left:$b$] (b) at (-2.4,-0.9);
        \fill (b) circle (1pt);

        \coordinate[label=right:$a$] (a) at (1.4,-0.9);
        \fill (a) circle (1pt);

        \coordinate[label=above:{$e_1(a)$}] (ap) at (0.5,0.1);
        \fill (ap) circle (1pt);

        \coordinate[label=above:$e_2(a)$] (ap2) at (1.5,0.2);
        \fill (ap2) circle (1pt);

        \coordinate[label=above:$e_3(a)$] (ap3) at (2.5,0.1);
        \fill (ap3) circle (1pt);

        \path[] (a) edge[draw=gray, dashed] node [sloped, allow upside down, inner sep=1pt, fill=white]{\small$\sqsubseteq$} (ap);
        \path[] (a) edge[draw=gray, dashed] node [sloped, allow upside down, inner sep=1pt, fill=white]{\small$\sqsubseteq$} (ap2);
        \path[] (a) edge[draw=gray, dashed] node [sloped, allow upside down, inner sep=1pt, fill=white]{\small$\sqsubseteq$} (ap3);

        \draw[-latex] (b) to[bend right=15] node[below]{\small$\beta$} ($(a) - (1pt, 0pt)$);

        \draw[-latex] (a) to[bend right=15] node[above]{\small$\gamma$} (-2.0, -0.2);

        \draw[-latex] (ap) to[bend right=15] node[above]{\small$\gamma$} (-1.8, 0.6);

        \draw[-latex, thick] (3.2, -1.3) to node[sloped, below]{more general} (3.2, 1.3);

    \end{tikzpicture}

%% file: pics/tikz/gen_tree.tex
    \begin{tikzpicture}
      \tikzstyle{basic}=[thick,align=left, text width=3.9cm]
      \tikzstyle{taken}=[draw=green!70!black, basic]
      \tikzstyle{nottaken}=[draw=red!70!black, basic]

      \tikzstyle{myarrow}=[semithick, draw=black, -stealth]

      \node[basic, draw=black] (n1) {
        \small
            \textbf{Instructions:}\hfill(1)\\
            \begin{tabular}{@{\hskip 0pt}c@{~}l}
              1. & mnemonic: \texttt{mov}; mem: W\\
              2. & mnemonic: \texttt{add}; mem: R
            \end{tabular}\\
            \textbf{Aliasing:}\\
            \begin{tabular}{l}
              op 1 of insn 1 must alias\\ with op 2 of insn 2
            \end{tabular}
        };

      \node[taken, below right=0.3cm of n1.south] (n3) {
        \small
            \textbf{Instructions:}\hfill(3)\\
            \begin{tabular}{@{\hskip 0pt}c@{~}l}
              1. & mnemonic: \texttt{mov}; mem: W\\
              2. & mnemonic: \fbox{$\top$}; mem: R
            \end{tabular}\\
            \textbf{Aliasing:}\\
            \begin{tabular}{l}
              op 1 of insn 1 must alias\\ with op 2 of insn 2
            \end{tabular}
        };

      \node[nottaken, below left=0.3cm of n1.south] (n2) {
        \small
            \textbf{Instructions:}\hfill(2)\\
            \begin{tabular}{@{\hskip 0pt}c@{~}l}
              1. & mnemonic: \texttt{mov}; mem: W\\
              2. & mnemonic: \texttt{add}; mem: R
            \end{tabular}\\
            \textbf{Aliasing:}\\
            \begin{tabular}{l}
              \fbox{$\top$}
            \end{tabular}
        };

      \node[nottaken, below left=0.4cm of n3.south] (n5) {
        \small
            \textbf{Instructions:}\hfill(5)\\
            \begin{tabular}{@{\hskip 0pt}c@{~}l}
              1. & mnemonic: \texttt{mov}; mem: \fbox{$\top$}\\
              2. & mnemonic: $\top$; mem: R
            \end{tabular}\\
            \textbf{Aliasing:}\\
            \begin{tabular}{l}
              op 1 of insn 1 must alias\\ with op 2 of insn 2
            \end{tabular}
        };

      \node[nottaken, left=0.3cm of n5.west] (n4) {
        \small
            \textbf{Instructions:}\hfill(4)\\
            \begin{tabular}{@{\hskip 0pt}c@{~}l}
              1. & mnemonic: \texttt{mov}; mem: W\\
              2. & mnemonic: $\top$; mem: \fbox{$\top$}
            \end{tabular}\\
            \textbf{Aliasing:}\\
            \begin{tabular}{l}
              op 1 of insn 1 must alias\\ with op 2 of insn 2
            \end{tabular}
        };
      \node[taken, right=0.3cm of n5.east] (n6) {
        \small
            \textbf{Instructions:}\hfill(6)\\
            \begin{tabular}{@{\hskip 0pt}c@{~}l}
              1. & mnemonic: \fbox{$\top$}; mem: W\\
              2. & mnemonic: $\top$; mem: R
            \end{tabular}\\
            \textbf{Aliasing:}\\
            \begin{tabular}{l}
              op 1 of insn 1 must alias\\ with op 2 of insn 2
            \end{tabular}
        };

      \draw[myarrow] (n1) -- (n2);
      \draw[myarrow] (n1) -- (n3);
      \draw[myarrow] ($(n3.south west) + (0cm, 0.2cm)$) -- (n4.north east);
      \draw[myarrow] (n3) -- (n5);
      \draw[myarrow] (n3) -- (n6);

    \end{tikzpicture}

%% file: pics/tikz/abstraction_singleton.tex
    \begin{tikzpicture}
      \small
      \node[] (t) {$\top$};
      \node[below left=0.5 and 0.3 of t] (n1) {$x_1$};
      \node[below=0.5 of t] (n2) {$x_2$};
      \node[below right=0.5 and 0.3 of t] (n3) {\dots};

      \draw[] (n1) -- (t);
      \draw[] (n2) -- (t);
      \draw[] (n3) -- (t);
    \end{tikzpicture}

%% file: pics/tikz/abstraction_edits.tex
    \begin{tikzpicture}
      \small
      \node[] (t) {$\top$};
      \node[below left=0.15 and 0.7 of t] (n1) {$x_1 + K \,\mathit{edits}$};
      \node[below=0.15 of t] (n2) {$x_2 + K \,\mathit{edits}$};
      \node[below right=0.15 and 0.7 of t] (n3) {\dots};

      \node[below=0.15 of n1] (n11) {\dots};
      \node[below=0.15 of n2] (n22) {\dots};

      \node[below=0.15 of n11] (n111) {$x_1 + 0 \,\mathit{edits}$};
      \node[below=0.15 of n22] (n222) {$x_2 + 0 \,\mathit{edits}$};

      \draw[] (n111) -- (n11) -- (n1) -- (t);
      \draw[] (n222) -- (n22) -- (n2) -- (t);
      \draw[] (n3) -- (t);
    \end{tikzpicture}

%% file: pics/tikz/abstraction_logsizes.tex
    \begin{tikzpicture}
      \small
      \node[] (t) {$\top$};
      \node[below=0.15 of t] (n1) {$\abs{x}< 2^K$};

      \node[below=0.15 of n1] (n11) {\dots};

      \node[below=0.15 of n11] (n111) {$\abs{x} < 1$};

      \draw[] (n111) -- (n11) -- (n1) -- (t);
    \end{tikzpicture}

%% file: pics/tikz/abstraction_subset_or_dn.tex
    \begin{tikzpicture}
      \small
      \node[] (t) {$\top$};

      \node[below left=0.00 and 0.2 of t] (n1) {$\{\,\}$};

      \node[below right=0.00 and 0.2 of t] (none) {\textsf{DefNone}};

      \node[below left=0.1 and 0.05 of n1] (n2) {$\{x_1\}$};
      \node[below right=0.1 and 0.05 of n1] (n3) {$\{x_2\}$};
      \node[right=0.3 of n3] (n4) {$\{\dots\}$};

      \node[below=0.7 of n1] (n5) {$\{x_1, x_2\}$};

      \node[right=0.2 of n5] (n6) {$\{x_2, \dots\}$};

      \node[left=0.2 of n5] (n55) {$\{x_1, \dots\}$};


      \draw[] (n1) -- (t);
      \draw[] (none) -- (t);
      \draw[] (n2) -- (n1);
      \draw[] (n3) -- (n1);
      \draw[] (n4) -- (n1);
      \draw[] (n5) -- (n2);
      \draw[] (n5) -- (n3);
      \draw[] (n6) -- (n3);
      \draw[] (n55) -- (n2);

    \end{tikzpicture}

%% file: tex/eval.tex
\section{Evaluation}
\label{sec:eval}

The main goal of \approach is to provide insights into the basic block throughput predictors under investigation.
Since this goal is not easily quantified, we evaluate \approach in two parts:
a general investigation of how inconsistencies are generalized (\autoref{ssec:effectiveness}) and a number of detailed case studies to give examples of actual insights gained (\autoref{sec:case-studies}).

\subsection{Considered Tools}

We compare a broad range of throughput predictors:

\iaca~\cite{iaca} is a closed-source tool provided by Intel to estimate the performance of basic blocks on their microarchitectures.
In April 2019, Intel announced that \iaca has reached its end of life.
We use the last released version, 3.0.

\llvmmca~\cite{llvmmca}, \osaca~\cite{laukemann18osaca}, and \uica~\cite{abel21uica} are open source basic block throughput predictors.
Their processor models are constructed by hand from documentation, contributed by hardware vendors, or inferred from measurements.
\llvmmca uses the instruction scheduling models of the \llvm compiler infrastructure~\cite{lattner04}.
If not stated otherwise, we use release 13 of \llvmmca, version 0.4.6 of \osaca, and commit \texttt{71f2eb6} from \uica's GitHub repository.\footnote{\url{https://github.com/andreas-abel/uiCA}}

In contrast to this, \ithemal~\cite{mendis19ithemal} and \difftune~\cite{renda2020difftune} infer their models through machine learning.
\ithemal predicts throughputs through an LSTM-based neural network that is trained on throughput measurements for a set of basic blocks.
We use their provided model that was trained on basic blocks from the BHive~\cite{chen19} data set (commits \texttt{47a5734} and \texttt{87c2468} of the corresponding GitHub repositories\footnote{\url{https://github.com/ithemal/Ithemal} and \url{https://github.com/ithemal/Ithemal-models}}).
\difftune is a modified version of \llvmmca where parameters of the processor model are replaced with learned ones.
We use the parameters that the authors provide, which were obtained through surrogate learning with an \ithemal-based model (commit \texttt{9992f69} in the GitHub repository\footnote{\url{https://github.com/ithemal/DiffTune}}).

We do not compare the MAQAO Code Quality Analyzer~\cite{rubial14cqa} in this evaluation as it requires a loop as input for its throughput prediction, which not all of the other tools support.\footnote{Our \approach implementation nevertheless supports tools like MAQAO, with an option to wrap each basic block in a loop when it is given as input to the tools.}

\subsection{\approach Parameters}

We use the following parameters for \approach:
\begin{itemize}[leftmargin=*]
  \item Threshold that the relative difference between to predictions must exceed to be considered interesting: 0.5
  \item Number of samples to check whether an abstract block is interesting: 100
  \item Maximal length of sampled basic blocks for discovery: 5 instructions
  \item Number of randomized generalizations per basic block: 5
\end{itemize}

In preliminary experiments, we found that variations in the latter three parameters do not affect \approach's results substantially in terms of the metrics presented in this section.
Only if they are selected widely out of range (\eg, only using very few samples to check for interestingness or only investigating very short basic blocks), the performance declines.
The threshold of the interestingness metric is of more relevance since it determines what inconsistencies are found.
The selected value is relatively large, which causes \approach to focus on substantial output differences.
We found such inconsistencies to be more likely to hint at conceptual differences like the handling of memory dependencies (\cf \autoref{ssec:cs-mempdeps}).

We extract the instruction schemes used for sampling from uops.info~\cite{abel19uopsinfo}.
For the evaluation, we exclude instruction schemes if they satisfy any of the following conditions:
\begin{itemize}[leftmargin=*]
  \item They are in a SIMD or FP extension other than AVX1\&2.
  \item They are not measured by uops.info.\footnote{This proxy criterion is intended to exclude instructions that are not supported in the microarchitecture, \eg, because they are from outdated ISA extensions.
    }
  \item They affect control flow.
  \item They need to be executed in privileged mode.
\end{itemize}
This leaves us with 2940 instruction schemes.
For each campaign, we further exclude all instructions that are not supported by one of the tools under investigation.\footnote{We consider an instruction supported by a tool if the tool gives a non-zero prediction for a basic block consisting of only the instruction.}
We configure the throughput predictors to assume the Intel Haswell microarchitecture since it is the only one supported by all considered tools.

The \approach campaigns ran on a system with an Intel Core i9-10900K processor (10 cores, 20 threads, 3.7\,GHz) and 64\,GB of RAM.
Running the predictors to evaluate the interestingness of basic blocks, which constitutes most of the execution time, is performed with 20 concurrent threads.

\begin{table*}
  \caption{\approach campaigns to find 150 inconsistencies, with metrics on how many basic blocks from \autoref{fig:intro_motivation} they explain, ordered by the percentage of interesting basic blocks subsumed.
    }
    \label{tab:coverage}
    \centering \small
  \hspace*{-0.8cm}
\begin{tabular}{r|c|c|c|c|c|c|c|c|c|c|}
   & \rot{\iaca,\,\osaca} & \rot{\llvmmca\,13,\,9} & \rot{\iaca,\,\uica} & \rot{\osaca,\,\uica} & \rot{\llvmmca,\,\uica} & \rot{\llvmmca,\,\osaca} & \rot{\iaca,\,\llvmmca} & \rot{\difftune,\,\osaca} & \rot{\iaca,\,\ithemal} & \rot{\ithemal,\,\osaca}\\
  \hline
   BBs interesting & 26\% & 23\% & 34\% & 32\% & 31\% & 34\% & 19\% & 52\% & 57\% & 47\%\\
  int. BBs covered & 97\% & 97\% & 91\% & 85\% & 83\% & 77\% & 74\% & 69\% & 68\% & 66\%\\
  \dots by top 10  & 68\% & 92\% & 82\% & 53\% & 72\% & 55\% & 70\% & 34\% & 54\% & 33\%\\
  run time (h:m)   & 6:34 & 0:32 & 1:35 & 6:59 & 1:19 & 5:28 & 0:38 & 9:29 & 4:34 & 8:13\\
  \hline
\end{tabular}
\begin{tabular}{r|c|c|c|c|c|c|}
   & \rot{\difftune,\,\llvmmca} & \rot{\difftune,\,\iaca} & \rot{\ithemal,\,\llvmmca} & \rot{\difftune,\,\uica} & \rot{\difftune,\,\ithemal} & \rot{\ithemal,\,\uica}\\
  \hline
   BBs interesting & 46\% & 52\% & 50\% & 40\% & 46\% & 30\%\\
  int. BBs covered & 63\% & 62\% & 62\% & 59\% & 57\% & 38\%\\
  \dots by top 10  & 31\% & 32\% & 49\% & 34\% & 29\% & 16\%\\
  run time (h:m)   & 5:55 & 5:54 & 5:05 & 6:25 & 10:02 & 5:15\\
  \hline
\end{tabular}

\end{table*}
\subsection{Generalization of Inconsistencies}
\label{ssec:effectiveness}

\autoref{fig:intro_motivation} in \autoref{sec:intro} demonstrates that we can find a large number of inconsistencies among the tools through random testing; enough that investigating them all by hand would be infeasible.
This section evaluates how \approach summarizes these inconsistencies.

The evaluation is based on the same data as \autoref{fig:intro_motivation}: a test set of 10,000 randomly sampled basic blocks consisting of 4 instructions each.
We sample these as described in \autoref{ssec:sampling} from an abstract block with 4 instructions and no constraints.
We ran \approach for each pair of tools until around 150 discoveries were found.
\autoref{tab:coverage} contains a column for each \approach campaign.\footnote{For brevity, we only include a comparison of \llvmmca version 9 and the current \llvmmca version 13.}
The first line repeats the data from \autoref{fig:intro_motivation}: the percentage of basic blocks in the test set that are interesting, \ie, for which the relative difference of the predictions exceeds 50\% of their average.

The second line shows the percentage of the set of interesting basic blocks from the test set that are subsumed (\cf \autoref{ssec:subsumption}) by an \approach discovery.
At 74\% to 97\%, these ratios are very high for comparisons of \iaca, \llvmmca, \uica, and \osaca.
This indicates that \approach inferred general descriptions of the differences between these tools.

The third line further demonstrates that \approach effectively condenses the inconsistent basic blocks for manual inspection:
it gives the ratio of interesting basic blocks in the test set that are subsumed by a subset of only ten discoveries of the \approach campaign.
In eight of the campaigns, these numbers were higher than 50\%, meaning that in these cases only ten of \approach's discoveries are sufficient to plausibly explain more than half of the inconsistently predicted basic blocks.
In every case except for the \ithemal/\uica campaign, ten of the \approach discoveries subsume more than 1000 inconsistent basic blocks from the dataset, ranging up to 3060 subsumed inconsistencies in the \iaca/\ithemal campaign.
The discovery subsets for this metric were computed with the strategy to maximize the number of subsumed basic blocks presented in \autoref{ssec:discoveryranking-subset}, applied to the interesting basic blocks in the test set.

The time required to find these discoveries, as displayed in the last line, mainly depends on how fast the tools produce their predictions.
While not the focus of this work, we observe that in this setup, \iaca, \llvmmca, and \uica were considerably faster than \osaca, \ithemal, and \difftune.

The campaigns that include \ithemal and \difftune still cover a substantial number of inconsistencies, but \approach finds less potential for generalization here than in the other campaigns.
We can identify reasons for this observation from the results for these campaigns:
\approach's generalizations terminate early in several instances where these tools produce results that run counter to common expectations.

For example, \ithemal produces different results for basic blocks that only differ in the specific register that they use, as can be seen in its predictions for basic blocks consisting of a single \enquote{rotate left} operation:
\begin{center}
\begin{tabular}{l|c|c}
  \textit{Basic Block} & \texttt{rol r12, cl} & \texttt{rol r10, cl}\\\hline
  \textit{Predicted Cycles} & 0.35 & 1.01
\end{tabular}
\end{center}
All other tools predict equal throughputs for these blocks.

\approach groups instructions by instruction schemes, \ie, a form that abstracts from the specific operands of the instruction.
It therefore does not generalize inconsistencies that are not independent of the concrete registers used.
Most throughput predictors share this notion and do not change their prediction if, \eg, operand registers in the basic block are replaced (while preserving dependencies).
This assumption is evidently not enforced in \ithemal's neural network.

\approach's results demonstrate this issue and therefore justify the insight that \ithemal might, \eg, benefit from training data where basic blocks are included multiple times with different but semantically equivalent register allocations.
Since the measured throughputs for these would be the same, the neural network might learn to abstract from the specific register used.

\begin{figure*}
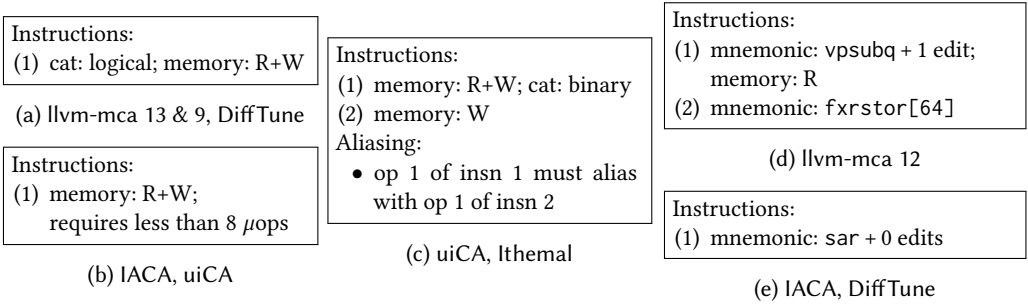

  \begin{minipage}{.31\textwidth}
    \begin{subfigure}[c]{\textwidth}
        \centering
        \fbox{\small
          \begin{minipage}{0.9\textwidth}
            \noindent
            Instructions:
            \begin{enumerate}[leftmargin=0.5cm, topsep=0.0cm]
              \item cat: logical; memory: R+W
            \end{enumerate}
          \end{minipage}
        }
      \caption{\llvmmca 13 \& 9, \difftune}
      \label{fig:cs-memdep-llvmmca-difftune}
    \end{subfigure}\\[4pt]

    \begin{subfigure}[c]{\textwidth}
        \centering
        \fbox{\small
          \begin{minipage}{0.9\textwidth}
            \noindent
            Instructions:
            \begin{enumerate}[leftmargin=0.5cm, topsep=0.0cm]
              \item memory: R+W;\\ requires less than 8 $\mu$ops
            \end{enumerate}
          \end{minipage}
        }
        \caption{\iaca, \uica}
        \label{fig:cs-memdep-uica-iaca}
    \end{subfigure}
  \end{minipage}%
  \begin{minipage}{.32\textwidth}
      \begin{subfigure}{\textwidth}
        \centering
        \fbox{\small
          \begin{minipage}{0.9\textwidth}
            \noindent
            Instructions:
            \begin{enumerate}[leftmargin=0.5cm, topsep=0.0cm]
              \item memory: R+W; cat: binary
              \item memory: W
            \end{enumerate}

            \noindent
            Aliasing:
            \begin{itemize}[leftmargin=0.5cm, topsep=0.0cm]
              \item op 1 of insn 1 must alias with op 1 of insn 2
            \end{itemize}
          \end{minipage}
        }
        \caption{\uica, \ithemal}
        \label{fig:cs-memdep-uica-ithemal}
      \end{subfigure}
  \end{minipage}%
  \begin{minipage}{.36\textwidth}
    \begin{subfigure}[c]{\textwidth}
      \centering
      \fbox{\small
        \begin{minipage}{0.9\textwidth}
          \noindent
          Instructions:
          \begin{enumerate}[leftmargin=0.5cm, topsep=0.0cm]
            \item mnemonic: \texttt{vpsubq}\,+\,1\:edit;\\memory: R
            \item mnemonic: \texttt{fxrstor[64]}
          \end{enumerate}
        \end{minipage}
      }
    \caption{\llvmmca 12}
    \label{fig:cs-fxrstor}
    \end{subfigure}\\[4pt]

    \begin{subfigure}[c]{\textwidth}
      \centering
      \fbox{\small
        \begin{minipage}{0.9\textwidth}
          \noindent
          Instructions:
          \begin{enumerate}[leftmargin=0.5cm, topsep=0.0cm]
            \item mnemonic: \texttt{sar}\,+\,0\:edits
          \end{enumerate}
        \end{minipage}
      }
    \caption{\iaca, \difftune}
    \label{fig:cs-difftune-shifts}
    \end{subfigure}
  \end{minipage}
  \caption{Abstract blocks causing inconsistent behavior found by \approach. Feature abstractions are summarized for brevity.}
  \label{fig:case-studies}
  \Description{
    The figure shows 5 abstract blocks:
    (a) for llvm-mca 13 and 9 as well as DiffTune: A single abstract instruction requiring the category "logical" and that memory is read and written.
    (b) for IACA and uiCA: A single abstract instruction requiring that memory is read and written and that less than 8 micro ops are used to execute represented instructions.
    (c) for uiCA and Ithemal: Two abstract instructions: The first requires that memory is read and written and the category "binary"; the second one that memory is written. The aliasing constraint requires that the read/written operand of instruction 1 is written by instruction 2.
    (d) for llvm-mca 12: Two abstract instructions: the first requires that memory is read and a mnemonic with edit distance less than or equal to 1 from "vpsubq"; the second requires an "fxrstor" instruction.
    (e) for IACA and DiffTune: A single abstract instruction requiring the mnemonic to be exactly "sar".
  }
\end{figure*}

\difftune learns parameters for \llvmmca and can therefore not produce different predictions based on the specific operands of the instructions as \ithemal does.
We can however observe that instructions that are very similar are predicted differently by \difftune.
For example, \approach finds that the abstract block in \autoref{fig:cs-difftune-shifts} represents an inconsistency between \difftune and \iaca.
This abstract block covers arithmetic right shift instructions, which, as the witnessing experiments in \approach's generalization decision tree show, \difftune predicts slower than \iaca if they use memory and faster if they do not use memory.
However, this discovery also indicates that instructions with a mnemonic that is only slightly different, like the logical right shift operations \texttt{shr}, are not predicted inconsistently.
From the experiments that reject the expansion to a mnemonic edit distance of 1, we can see that \difftune gives different predictions for \texttt{shr} and \texttt{sar} instructions, in contrast to most other tools.
This different treatment of similar instructions invites for a closer inspection, but it restricts \approach's generalizations.

In summary, we observe that \approach's generalization is very effective for the majority of considered tools.
Where generalization is not as effective, the results are nevertheless insightful and point to concrete problems.

\section{Case Studies}
\label{sec:case-studies}
The previous section shows that \approach is able to summarize thousands of inconsistencies between throughput predictors by a small number of abstract blocks.
For \difftune and \ithemal, it also presents first lessons learned from \approach's results.
We further investigated \approach's discoveries and found several kinds of insights, for which we present examples in the following:
\begin{itemize}[leftmargin=*]
  \item They uncover different assumptions in the tools that can lead to dramatically different predictions. (\autoref{ssec:cs-mempdeps})
  \item They find newly introduced regression bugs in subsequent versions of the same tool (\autoref{ssec:cs-mempdeps}) as well as long-existing bugs (\autoref{ssec:cs-fxrstor}).
  \item \approach can characterize a variety of inaccuracies in \llvmmca's model for the AMD Zen+ microarchitecture, as well as an unusual quirk in the microarchitecture itself. (\autoref{ssec:cs-llvmmca})
\end{itemize}

\subsection{Memory Dependencies}
\label{ssec:cs-mempdeps}

Data dependencies through memory operands are a challenge for basic block throughput predictors.
If subsequent writes to and reads from memory refer to the same location, the write needs to be completed before the read.\footnote{More specifically, the written value needs at least to be computed and put into a store buffer, from which it can be forwarded to subsequent reads.}
If they access disjoint memory locations, they can execute independently.
However, which of the cases applies may not be obvious or depend on the inputs.
\approach's results show that the tools handle these cases quite differently.

\begin{table}
  \caption{Predictions for the cycles required to execute basic blocks that differ in their memory dependencies.}
  \label{tab:deps-predictions}
  \centering \small
\hspace*{-0.5cm} 
\begin{tabular}{r|ccccccccc}
  \setlength{\tabcolsep}{1pt}
  Basic Block& \rotless{\uica} & \rotless{\osaca} & \rotless{\iaca} & \rotless{\llvmmca 13} & \rotless{\llvmmca 13 alias} & \rotless{\llvmmca 9} & \rotless{\difftune} & \rotless{\ithemal}\\\hline
  \texttt{add [rcx+16],rbx; add [\textbf{rcx+16}],rbx}  & 12.0 & 12.0& 2.0 & 2.1 & 14.0 & 14.0 & 14.1 & 5.9 \\
  \texttt{add [rcx+16],rbx; add [\textbf{rcx+128}],rbx} & 6.0  & 6.0 & 2.0 & 2.1 & 14.0 & 14.0 & 14.1 & 5.9 \\
  \texttt{add [rcx+16],rbx; add [\textbf{rdx+16}],rbx}  & 6.0  & 6.0 & 2.0 & 2.1 & 14.0 & 14.0 & 14.1 & 6.0
\end{tabular}
\end{table}

\autoref{tab:deps-predictions} shows how the throughput predictors handle memory dependencies on three example basic blocks: one with a guaranteed memory dependency (first line), one with independent instructions (second line), and one where the instructions may be independent, given suitable register values (third line).
We see three plausible throughput prediction results for executing such basic blocks in a loop:
\begin{itemize}[leftmargin=*]
  \item Two cycles, if there are no dependencies through memory and each instruction uses the processor's store unit for one cycle.
  \item Six cycles, if there is no memory dependency between the two instructions, but each instruction depends on its own result from the previous iteration. They then form two dependency chains with a latency of 6 cycles, which can be executed in parallel.
  \item Around 12 cycles, if all memory accesses depend on each other, forming a single large dependency chain.
\end{itemize}
\llvmmca in its default setting~\cite{llvmmca-man} and \iaca\footnote{as previously noted by \citet{abel19uopsinfo}} assume the first case.
\approach shows that for \llvmmca in the outdated version 9, this was not the case:
it discovers, \eg, that the abstract block in \autoref{fig:cs-memdep-llvmmca-difftune} represents an inconsistency between \llvmmca version 9 and 13.
The older version was affected by a bug that led to predictions as if all memory accesses aliased.
While this bug has been manually discovered and fixed in the past, \approach automatically finds this regression with a minimal example for reproducing the bug.

We also find that \difftune's learned parameters for \llvmmca attempt to bypass this assumption:
\approach finds the same abstract block in \autoref{fig:cs-memdep-llvmmca-difftune} in the campaign for \llvmmca 13 and \difftune.
The example basic blocks for this discovery show that \difftune also predicts throughputs as if all accesses were dependent on each other.
Regarding this, \citet{renda2020difftune} remark in their evaluation that \difftune learned a ``degenerately high'' latency for instructions that read and write memory from the same location.
\llvmmca also provides an override switch to assume that all memory accesses alias, which leads to results similar to \difftune's.

\approach discoveries like the one in \autoref{fig:cs-memdep-uica-iaca} indicate that \uica, \osaca, and \ithemal do not share \iaca's assumption that no memory operations alias.
These three tools recognize the data dependency formed by a single instruction that reads \emph{and} writes with itself, therefore \approach reports no discovery like those in \autoref{fig:cs-memdep-llvmmca-difftune} and \ref{fig:cs-memdep-uica-iaca} between them.
However, the abstract block displayed in \autoref{fig:cs-memdep-uica-ithemal} allows us to identify the difference in line 1 of \autoref{tab:deps-predictions}:
\uica and \osaca rightly assume that the two memory locations alias if they are identical.
\approach provides the basic block treated in line 1 of the table for this abstract block.

For a user of these tools, this discrepancy can significantly affect the outcome:
if the memory operands in the application alias in a non-obvious way, the observed cycles would exceed the results of \uica and \osaca by a factor of two, and those of \iaca and the default setting of \llvmmca by a factor of 6.

Since this inconsistency affects a large number instruction schemes, corresponding discoveries were easy to find in \approach's report with discoveries ranked by their generality (\cf \autoref{ssec:discoveryranking-generality}).

\subsection{FXRSTOR Crash in \llvmmca}
\label{ssec:cs-fxrstor}

To uncover crashes, \approach can compare a single tool with itself.
\autoref{fig:cs-fxrstor} shows an abstract block we found when investigating \llvmmca.
This abstract block crashes the tool with an assertion, which \approach always counts as interesting.
\texttt{FXRSTOR} instructions, which restore the state of floating point control registers from memory, require the processor to execute a large number of micro operations.
If one of the resources that it accesses is also used by a different instruction nearby, \eg, a vector subtraction with a memory operand, a bug in \llvmmca is triggered.

For \llvm release 13, this bug has recently been reported and fixed independently of our research.\footnote{\url{http://bugs.llvm.org/PR50725}}
The report includes a large input with more than 300 instructions to trigger the bug.
\approach automatically discovered the issue and provides a minimal input of just two instructions.

This discovery appears prominently in \approach's results when the discoveries are ranked by their interestingness (\cf \autoref{ssec:discoveryranking-interestingness}) since crashes in a tool under investigation are reported as maximally interesting.

\subsection{Comparing \llvmmca to Measurements}
\label{ssec:cs-llvmmca}

\newcommand{\discoverywidth}{0.27\textwidth}

\begin{table*}
  \caption{Abstract blocks capturing inconsistent behavior found by \approach in \llvmmca/\nanobench campaigns on the AMD Zen+ microarchitecture.
    The descriptions are not generated by \approach. The \enquote{Resulting Cycles} column displays the number of cycles predicted by \llvmmca (mca) and the cycles measured by \nanobench (nb) for the basic block in the preceding column.}
  \label{fig:zenp_discoveries}
  \small
  \begin{tabular}{cccc||c}
    \toprule
     & Abstract Block & Example Basic Block & \parbox[c]{1.2cm}{\centering Resulting\\Cycles} & Simplified Description \\
    \midrule
    \newtag{(A)}{zenp-discovery:ucode} &
      {
        \begin{minipage}{\discoverywidth}
          \noindent
          Instructions:
          \begin{enumerate}[leftmargin=0.5cm, topsep=0.0cm]
            \item opschemes: $\{\verb|R:flag_df|\}$
          \end{enumerate}
        \end{minipage}
      } &
      {
        \footnotesize
        \texttt{cmpsq}
      } &
      \parbox[l]{1.2cm}{mca:\hfill 100\\nb:\hfill 3.0} &
      \multirow{3}{*}[-2ex]{
        \parbox[l]{3.3cm}{\llvmmca models complex instructions, certain shifts, and horizontal vector operations very pessimistically.}
      }\\
    \cmidrule{1-4}
    \newtag{(B)}{zenp-discovery:shld} &
      {
        \begin{minipage}{\discoverywidth}
          \noindent
          Instructions:
          \begin{enumerate}[leftmargin=0.5cm, topsep=0.0cm]
            \item opschemes: $\{\verb|R:cl|\}$\\isa-set: \texttt{I386}
          \end{enumerate}
        \end{minipage}
      } &
      {
        \footnotesize
        \texttt{shld r11,\,rdx,\,cl}
      } &
      \parbox[l]{1.2cm}{mca:\hfill 100\\nb:\hfill 3.0} & \\
    \cmidrule{1-4}
    \newtag{(C)}{zenp-discovery:horiz} &
      {
        \begin{minipage}{\discoverywidth}
          \noindent
          Instructions:
          \begin{enumerate}[leftmargin=0.5cm, topsep=0.0cm]
            \item mnemonic: \texttt{haddpd}\\\,+\,3\,edits; category: SSE3
          \end{enumerate}
        \end{minipage}
      } &
      {
        \footnotesize
        \texttt{hsubpd\;xmm15,\,xmm12}
      } &
    \parbox[l]{1.2cm}{mca:\hfill 100\\nb:\hfill 6.5} & \\

    \midrule
    \newtag{(D)}{zenp-discovery:bitscan} &
        {
          \begin{minipage}{\discoverywidth}
            \noindent
            Instructions:
            \begin{enumerate}[leftmargin=0.5cm, topsep=0.0cm]
              \item mnemonic: \texttt{bsf}\,+\,1\,edit\\
                    opschemes: $\{\verb|W:GPR:64|\}$
            \end{enumerate}
            \noindent
            Aliasing:
            \begin{itemize}[leftmargin=0.5cm, topsep=0.0cm]
              \item op 1 of insn 1 must not alias with op 2 of insn 1
            \end{itemize}
          \end{minipage}
        } &
      {
        \footnotesize
          \texttt{bsr rcx,\,r11}
      } &
    \parbox[l]{1.2cm}{mca:\hfill 0.3\\nb:\hfill 4.0} &
    \parbox[l]{3.3cm}{\llvm's model for bit-scan instructions implies a wrong throughput.} \\
    \midrule
    \newtag{(E)}{zenp-discovery:loads} &
        {
          \begin{minipage}{\discoverywidth}
            \noindent
            Instructions:
            \begin{enumerate}[leftmargin=0.5cm, topsep=0.0cm]
              \item mnemonic: \texttt{add}\,+\,2\,edits\\
                opschemes: $\{\verb|RW:GPR:64|\}$\\
                memory: \textsf{DefNone}
              \item mnemonic: \texttt{and}\,+\,3\,edits\\
                opschemes: $\{\verb|RW:GPR:64|\}$\\
                memory: R
            \end{enumerate}
            \noindent
            Aliasing:
            \begin{itemize}[leftmargin=0.5cm, topsep=0.0cm]
              \item op 1 of insn 1 must alias with op 1 of insn 2
            \end{itemize}
          \end{minipage}
        } &
      {
        \footnotesize
        \parbox[l]{2.8cm}{
          \texttt{add\;r8,\,0x2a}\\
          \texttt{adc\;r8,\,qword\,ptr\,[r14]}
        }
      } &
      \parbox[l]{1.2cm}{mca:\hfill 5.03\\nb:\hfill 2.0} &
      \parbox[l]{3.3cm}{\llvmmca misses that memory loads can start before other operands are available.} \\
    \midrule
    \newtag{(F)}{zenp-discovery:zeroshift} &
        {
          \begin{minipage}{\discoverywidth}
            \noindent
            Instructions:
            \begin{enumerate}[leftmargin=0.5cm, topsep=0.0cm]
              \item mnemonic:\,\texttt{shl}\,+\,2\,edits\\
                    opschemes:\,$\{\verb|0x0|\}$\\
                    isa-set:\,I186
              \item opschemes:\,$\{\verb|R:flag_of|\}$
            \end{enumerate}
          \end{minipage}
        } &
        \parbox[l]{2.2cm} {
          \footnotesize
          \texttt{shl r9w,\,0x0}\\
          \texttt{setno r11b}
        } &
      \parbox[l]{1.2cm}{mca:\hfill 1.04\\nb:\hfill 25.7} &
      \parbox[l]{3.3cm}{Reading flags after a shift by 0 incurs a penalty on Zen+.} \\
    \bottomrule
  \end{tabular}
\end{table*}

\approach has little requirements on the tools under investigation:
they only need to produce a throughput estimate for a given basic block.
Consequently, we can also use \approach to compare a throughput predictor to a tool that runs input basic blocks as microbenchmarks on the actual hardware.

In this case study, we apply \approach to compare the predictions of \llvmmca's model for the AMD Zen/Zen+ microarchitecture to microbenchmarks performed with \nanobench~\cite{abel20nanobench} on an AMD Ryzen 5 2600X processor.
For these discoveries, we configured \approach to consider an abstract block interesting if the absolute difference between measured and predicted throughput of all of 50 sampled basic blocks is at least 2 cycles.
This ensures that we can identify the more subtle inconsistencies in the results.
We further restrict the instructions considered by \approach for sampling and generalization:
we exclude instruction schemes that read \emph{and} write memory to avoid discovering more variants of the problem described in \autoref{ssec:cs-mempdeps}.

It is important to note that the \nanobench measurements are just another tool under investigation, not a ground truth.
We configure \nanobench to group 10 instances of the measured instructions in a loop body that is executed 100 times while the passing processor cycles are measured with a hardware performance counter after 10 warm-up iterations.
We use defaults for the remaining settings of \nanobench, which entails among other things that most registers are initialized with arbitrary values (except for those used as memory addresses).
These assumptions affect the measured cycles, rendering the measurements unsuitable for use as definitive ground truth.
The strength of \approach's differential testing perspective is that neither tool needs to be assumed as \enquote{correct} to obtain interesting insights.

\autoref{fig:zenp_discoveries} shows the selected \approach discoveries that we discuss in the following.
We refer to the discoveries by the identifier in the first column.
The second column contains the abstract block reported as discovery by \approach.
With the generalization decision tree and the corresponding evaluated basic blocks, \approach provides more additional information than we can present here.
We therefore instead only show one example basic block sampled from the abstract block and the results of \nanobench (nb) and \llvmmca (mca) for it in the third and fourth columns.
In the last column, we annotate a short summary of the problem characterized by the discovery.

\paragraph{Microcoded Instructions}

When ranking \approach discoveries by their interestingness (\cf \autoref{ssec:discoveryranking-interestingness}), the ones that stand out the most are those concerning instructions that \llvmmca predicts to require 100 cycles to execute.
This mainly affects microcoded instructions, \eg, the string operations, which commonly read the direction flag register \texttt{df}, summarized by discovery \ref{zenp-discovery:ucode}.
When the processor's instruction decoder encounters such instructions, it produces a (potentially large and/or varying) number of \uops that need to be executed by the processor's functional units.
\llvm's Zen+ scheduling model (and consequently \llvmmca) handles most such instructions in a coarse way that just assigns them a latency of 100 cycles.

However, this strategy is also used for more unexpected instructions like certain bit shifts (discovery \ref{zenp-discovery:shld}) and horizontal vector operations (discovery \ref{zenp-discovery:horiz}).
While these modeling decisions are not per se bugs, they can make the Zen+ model of \llvmmca effectively unusable for any task that uses such instructions.
\llvm's issue tracker contains a report for this behavior that has been submitted independently of our work.\footnote{\url{https://github.com/llvm/llvm-project/issues/53242}}

\paragraph{Bit-Scan Instructions}
Discovery \ref{zenp-discovery:bitscan} shows an apparent bug in \llvm's Zen+ scheduling model for bit-scan instructions.\footnote{The bit-scan instructions \texttt{BSF}/\texttt{BSR} determine the index of the least/most significant bit set in their second operand and write it to the first operand.}
The measurements with \nanobench, as well as the instruction latency table provided by \citet{amd17}, show that a \texttt{BSR} instruction has a latency of 4 cycles and requires 4 cycles to be executed in a steady state (a throughput of 0.25 instructions per cycle).

The aliasing component of \approach's abstract block~\ref{zenp-discovery:bitscan} shows that \llvmmca predicts the latency consistently with measurements:
the must-not-alias constraint could not be dropped during generalization, which means that no inconsistency is found if the instruction forms a dependency chain with itself.
Without aliasing, \llvmmca underestimates the required execution time.
The scheduling model of \citet[l.\,235]{llvm-zen-model} provides an explanation: the affected instructions are modeled with a plausible latency, but they only use one of the architecture's four arithmetical/logical units.
Therefore, \llvmmca assumes that up to four independent instances of bit-scan instructions can be executed per cycle.
We reported this and four other results from similar \approach discoveries to the \llvm developers.
These reports show errors in \llvm's Zen+ scheduling model for a total of 72 of our instruction schemes.
The bugs were confirmed and fixed by the developers.

\paragraph{Inaccuracies in Load Operand Usage}
With discovery \ref{zenp-discovery:loads}, we learn that \llvmmca over-estimates the time required to execute instructions that depend on the result of a preceding instruction and load from memory.
The hardware is evidently able to issue a new instruction in every cycle for the corresponding example basic block; the load latency (4 cycles for L1 cache hits on Zen+) completely overlaps with the remaining computation.
\llvmmca's model does not account for this behavior: here, the loading instruction always starts executing with the instruction it depends on, causing the load latency to be visible.
This problem has also been independently reported in the \llvm issue tracker.\footnote{\url{https://github.com/llvm/llvm-project/issues/50899}}

\paragraph{A Microarchitectural Quirk}
\approach's results not only highlight oddities in prediction tools, they can also show unusual behavior in the processor under test.
The discovery \ref{zenp-discovery:zeroshift} shows how \approach automatically found a microarchitectural quirk of the Zen architectures that has been previously described by \citet{Abel20b}:
Bit shifts by zero (which invoke a special case where the flag registers are not updated\footnote{see, \eg, \url{https://www.felixcloutier.com/x86/sal:sar:shl:shr}}) cause a severe execution time penalty if they are followed by instructions that read the flag registers.
\llvmmca's model omits this unexpected corner case of AMD's Zen architectures.

\medskip

In all of the above cases, \approach automatically discovered an unexpected inconsistency and provided helpful insight with its generalization.
Such insights could otherwise only be gained through tedious manual effort.
The fact that we, additionally to finding new bugs, automatically rediscovered several previously reported problems in the \llvmmca predictions indicates that \approach finds problems that are relevant to the users of \llvmmca.

%% file: tex/relwork.tex
\section{Related Work}
\label{sec:relwork}

To the best of our knowledge, \approach is the first work to apply differential testing to microarchitectural code analyzers.
This section describes other approaches to evaluate such tools and contrasts \approach to previous work in differential testing.

\subsection{Testing Throughput Predictors}

Most of the available basic block throughput predictors come with an evaluation of their prediction accuracy.
A common approach to evaluating throughput predictors is to measure the relative error from and the correlation with execution time measurements on a chosen set of basic blocks.
This is done for \osaca~\cite{laukemann18osaca}, \ithemal~\cite{mendis19ithemal}, \difftune~\cite{renda2020difftune}, and \uica~\cite{abel21uica}.
They all use basic blocks that were extracted from the binaries of common benchmarks and open source programs whose throughput was measured using different methodologies.
Of particular note is \bhive~\cite{chen19}, which is used in the evaluations of \difftune and \uica.
It is an openly available set of such basic blocks with annotated measured throughputs for several Intel microarchitectures.
The evaluation of \uica identifies cases where assumptions made for the ground truth measurements affect which tool is ``more accurate'' than another, motivating our differential approach.

Evaluating the prediction accuracy on basic blocks from compiled programs is helpful when the expected use of the tools is on similar basic blocks.
However, such basic blocks are lacking when systematically exploring inconsistencies of the tools:
of the 2940 instruction schemes that we use in our evaluation, 2002 (\ie, 68\%) do not occur in any basic block of the \bhive data set and 525 (\ie, 18\%) of the instruction schemes are enough to represent 99\% of the \bhive basic blocks.
The \bhive benchmarks therefore leave a gap in the input space when testing throughput predictors that \approach intends to close.

\bhive also includes an approach to help developers identify problems with their throughput predictors.
They cluster basic blocks from the data set based on their use of execution units in the processor (\eg, ``vectorized code'' and ``code with mainly memory operations'').
If a tool performs particularly poor on a cluster of basic blocks, the developers can focus on improving support for the associated category.
This direction is however considerably less specific than the inconsistencies that \approach reports to the user.

A concurrently published work by \citet{abel2022difftune} investigates the prediction accuracy of \difftune, providing a very simple set of parameters for \llvmmca that outperform the learned \difftune parameters on the \bhive data set in terms of prediction accuracy.
These findings are consistent with the unexpected predictions we encountered in our \difftune campaigns (\autoref{ssec:effectiveness}).

\exegesis~\cite{exegesis,llvmexegesis,llvmexegesisman} is a project to validate \llvm's performance models and, consequently, \llvmmca.
For a given instruction scheme, \exegesis executes a microbenchmark on the target machine and measures its performance characteristics.
\exegesis can compare the measured performance to the corresponding information in \llvm's scheduling model.
In contrast to \approach, \exegesis does not generate experiments with multiple instructions to test their interactions and it is closely integrated with \llvm.
Comparisons with other predictors are therefore not supported.

Approaches that infer models for throughput predictors are also evaluated against existing ones on a measured ground truth:
\pmevo~\cite{pmevo} and \palmed~\cite{derumigny21palmed} both use basic blocks without data dependencies, whose throughput is bound by the processor's functional units.
For \palmed, the basic blocks mirror basic blocks observed in the binaries of benchmark suites (without the dependencies).
The basic blocks used for \pmevo are more similar to the ones we use here: they are randomly sampled in a way that avoids data dependencies.
The evaluation of \uopsinfo~\cite{abel19uopsinfo} points out some inconsistencies in \iaca, but focuses on the usage of resources for single instructions.

\subsection{Differential Testing}

Differential testing~\cite{mckeeman98} is commonly used to find bugs in tools where no ground truth is available.
There are general frameworks for differential testing tools like Nezha~\cite{petsios17nezha}, but they mainly focus on effectively exploring a sparse space of inconsistencies.
As the space of inconsistencies among basic block throughput predictors is not sparse, there is little benefit in using these frameworks.

\approach's use of minimization and abstraction can be seen as a form of triage in the usual nomenclature~\cite{manes19fuzzsurvey}.
The concepts and notations borrowed from abstract interpretation give us a way to systematically implement a generalized deduplication of inputs.

Previous research already used differential testing for other tools operating on machine code.
Several differential fuzzing approaches \cite{paleari2010,jay2018,woodruff2021differential} focus on instruction decoders.
However, these works differ from our setting in their goal and, consequently, in the inputs that they generate.
They generate bit sequences that are (or are close to) machine instructions, for which they check the results of a group of instruction decoders.
Since we aim to find inconsistencies in the throughput predictions, we only produce valid instruction sequences.
\citet{woodruff2021differential} note that they encounter large numbers of nearly identical discoveries that are difficult to deduplicate, making human analysis essential.
This mirrors our motivation to use abstraction to reduce the manual investigation effort for analyzing the discoveries.

Revizor~\cite{oleksenko21revizor} is a differential testing approach that also generates random instruction sequences.
They compare a CPU's behavior with that of a simulation that does not leak information to find side channel attacks.
In contrast to \approach, their instruction sequences include control flow.
They define a number of patterns on the dependencies between consecutive instructions that are similar to the constraints represented by our aliasing abstraction.
Revizor however uses these patterns only as a metric to control the size of the instruction sequences that they sample.
Since abstraction is central to \approach, we designed the basic block abstraction to cover more complex alias constraints as well as constraints on the involved instructions, which are beyond the scope of Revizor's patterns.

%% file: tex/conclusion.tex
\section{Possible Extensions}
\label{sec:extensions}

A strength of \approach is that the throughput predictors under investigation are treated as black boxes.
The resulting flexibility opens a range of further use cases for \approach with no or minor adjustments to the implementation:

\subsubsection*{Comparing Different Benchmarking Assumptions}

When benchmarking the execution time of basic blocks, tools like \nanobench~\citep{abel20nanobench} have to make assumptions on how the blocks should be executed.
For instance, they need to initialize registers and memory regions with specific values and choose whether basic blocks should be wrapped in a loop or concatenated sufficiently often.
If these choices are configurable (as with \nanobench), \approach can investigate the effect of different configuration decisions on the measurements.
From discovery \ref{zenp-discovery:zeroshift} in our \llvmmca case study (\autoref{ssec:cs-llvmmca}), we would, \eg, expect to find inconsistencies depending on whether the registers are initialized with 0 or not.

\subsubsection*{Comparing Measurements on Different Microarchitectures}

We have presented results for comparing pairs of throughput predictor tools (\autoref{ssec:effectiveness}) as well as for comparing a throughput predictor to measurements on the modeled hardware (\autoref{ssec:cs-llvmmca}).
A natural next step would be to compare measurements on two different hardware implementations of an instruction set architecture to each other.
This would allow us to investigate performance differences of subsequent generations of CPUs by the same manufacturer, or different trade-offs made by two manufacturers in competing CPU models.

\subsubsection*{Comparing Port Usage Models}

The \approach algorithm can also be applied to subcomponents of performance models that affect only individual aspects of basic block throughput prediction.
For instance, approaches like \uopsinfo~\citep{abel19uopsinfo}, \pmevo~\citep{pmevo}, and \palmed~\citep{derumigny21palmed} build models for how individual instructions use a CPU's execution resources.
These models are able to predict the throughput of basic blocks without data dependencies.

\approach could therefore investigate differences between the models produced by the individual approaches, as well as deviations between a model and measurements on the actual hardware.
For this application domain, the presented basic block abstraction should be adjusted such that only basic blocks with as few data dependencies as possible are sampled.
Consequently, the aliasing component of the basic block abstraction then does not capture meaningful information anymore and may be dropped.

Since these approaches infer their models from microbenchmarks, the results of \approach may be helpful to improve the models by characterizing classes of benchmarks that are missing.

\section{Conclusion}
\label{sec:conclusion}

State-of-the-art tools for basic block throughput prediction often do not agree in their results, for a variety of reasons.
To understand and improve them, we proposed \approach, a tool to differentially test basic block throughput predictors.
\approach uses notions from abstract interpretation to generalize inconsistencies in a systematic way.
Our evaluation shows that \approach can summarize thousands of inconsistencies in a few dozen descriptions that directly lead to high-level insights into the different behavior of the tools.

\approach further provides interesting points for future research: the core algorithms are independent of the application to throughput predictors and might therefore be of benefit in other domains of differential testing.

\section{Data Availability Statement}
This article is accompanied by an artifact \citep{anica-artifact}.
The artifact provides the implementation of the \approach algorithms used for our evaluation and case studies.
It also includes the results of the described \approach campaigns with a graphical user interface for inspection as well as means for reproducing them.
A development version of the \approach implementation is also available on Github at \url{https://github.com/cdl-saarland/AnICA}.

%% file: anica.bbl

\begin{thebibliography}{39}


\ifx \showCODEN    \undefined \def \showCODEN     #1{\unskip}     \fi
\ifx \showDOI      \undefined \def \showDOI       #1{#1}\fi
\ifx \showISBNx    \undefined \def \showISBNx     #1{\unskip}     \fi
\ifx \showISBNxiii \undefined \def \showISBNxiii  #1{\unskip}     \fi
\ifx \showISSN     \undefined \def \showISSN      #1{\unskip}     \fi
\ifx \showLCCN     \undefined \def \showLCCN      #1{\unskip}     \fi
\ifx \shownote     \undefined \def \shownote      #1{#1}          \fi
\ifx \showarticletitle \undefined \def \showarticletitle #1{#1}   \fi
\ifx \showURL      \undefined \def \showURL       {\relax}        \fi
\providecommand\bibfield[2]{#2}
\providecommand\bibinfo[2]{#2}
\providecommand\natexlab[1]{#1}
\providecommand\showeprint[2][]{arXiv:#2}

\bibitem[Abel(2020)]%
        {Abel20b}
\bibfield{author}{\bibinfo{person}{Andreas Abel}.}
  \bibinfo{year}{2020}\natexlab{}.
\newblock \emph{\bibinfo{title}{Automatic Generation of Models of
  Microarchitectures}}.
\newblock \bibinfo{thesistype}{Ph.\,D. Dissertation}.
  \bibinfo{school}{Universit\"at des Saarlandes}.
\newblock
\urldef\tempurl%
\url{https://d-nb.info/1212853466/34}
\showURL{%
\tempurl}


\bibitem[Abel(2022)]%
        {abel2022difftune}
\bibfield{author}{\bibinfo{person}{Andreas Abel}.}
  \bibinfo{year}{2022}\natexlab{}.
\newblock \showarticletitle{{DiffTune} Revisited: A Simple Baseline for
  Evaluating Learned llvm-mca Parameters}. In \bibinfo{booktitle}{\emph{Machine
  Learning for Computer Architecture and Systems 2022}}.
\newblock
\urldef\tempurl%
\url{https://openreview.net/forum?id=dw4evoj6AE}
\showURL{%
\tempurl}


\bibitem[Abel and Reineke(2019)]%
        {abel19uopsinfo}
\bibfield{author}{\bibinfo{person}{Andreas Abel} {and} \bibinfo{person}{Jan
  Reineke}.} \bibinfo{year}{2019}\natexlab{}.
\newblock \showarticletitle{uops.info: Characterizing Latency, Throughput, and
  Port Usage of Instructions on Intel Microarchitectures}. In
  \bibinfo{booktitle}{\emph{Proceedings of the Twenty-Fourth International
  Conference on Architectural Support for Programming Languages and Operating
  Systems}} (Providence, RI, USA) \emph{(\bibinfo{series}{ASPLOS '19})}.
  \bibinfo{publisher}{ACM}, \bibinfo{address}{New York, NY, USA},
  \bibinfo{pages}{673--686}.
\newblock
\showISBNx{978-1-4503-6240-5}
\urldef\tempurl%
\url{https://doi.org/10.1145/3297858.3304062}
\showDOI{\tempurl}


\bibitem[Abel and Reineke(2020)]%
        {abel20nanobench}
\bibfield{author}{\bibinfo{person}{Andreas Abel} {and} \bibinfo{person}{Jan
  Reineke}.} \bibinfo{year}{2020}\natexlab{}.
\newblock \showarticletitle{{nanoBench}: A Low-Overhead Tool for Running
  Microbenchmarks on x86 Systems}. In \bibinfo{booktitle}{\emph{2020 IEEE
  International Symposium on Performance Analysis of Systems and Software
  (ISPASS)}}. IEEE, \bibinfo{pages}{34--46}.
\newblock
\urldef\tempurl%
\url{https://doi.org/10.1109/ISPASS48437.2020.00014}
\showDOI{\tempurl}


\bibitem[Abel and Reineke(2022)]%
        {abel21uica}
\bibfield{author}{\bibinfo{person}{Andreas Abel} {and} \bibinfo{person}{Jan
  Reineke}.} \bibinfo{year}{2022}\natexlab{}.
\newblock \showarticletitle{{uiCA}: Accurate Throughput Prediction of Basic
  Blocks on Recent {Intel} Microarchitectures}. In
  \bibinfo{booktitle}{\emph{{ICS} '22: 2022 International Conference on
  Supercomputing, Virtual Event, June 28 - 30, 2022}},
  \bibfield{editor}{\bibinfo{person}{Lawrence Rauchwerger},
  \bibinfo{person}{Kirk~W. Cameron}, \bibinfo{person}{Dimitrios~S.
  Nikolopoulos}, {and} \bibinfo{person}{Dionisios~N. Pnevmatikatos}} (Eds.).
  \bibinfo{publisher}{{ACM}}, \bibinfo{pages}{33:1--33:14}.
\newblock
\urldef\tempurl%
\url{https://doi.org/10.1145/3524059.3532396}
\showDOI{\tempurl}


\bibitem[AMD(2017)]%
        {amd17}
\bibfield{author}{\bibinfo{person}{AMD}.} \bibinfo{year}{2017}\natexlab{}.
\newblock \bibinfo{booktitle}{\emph{Software Optimization Guide for AMD Family
  17h Processors}}.
\newblock \bibinfo{publisher}{AMD}.
\newblock


\bibitem[Binkert et~al\mbox{.}(2011)]%
        {binkert11gem5}
\bibfield{author}{\bibinfo{person}{Nathan~L. Binkert},
  \bibinfo{person}{Bradford~M. Beckmann}, \bibinfo{person}{Gabriel Black},
  \bibinfo{person}{Steven~K. Reinhardt}, \bibinfo{person}{Ali~G. Saidi},
  \bibinfo{person}{Arkaprava Basu}, \bibinfo{person}{Joel Hestness},
  \bibinfo{person}{Derek Hower}, \bibinfo{person}{Tushar Krishna},
  \bibinfo{person}{Somayeh Sardashti}, \bibinfo{person}{Rathijit Sen},
  \bibinfo{person}{Korey Sewell}, \bibinfo{person}{Muhammad Shoaib~Bin Altaf},
  \bibinfo{person}{Nilay Vaish}, \bibinfo{person}{Mark~D. Hill}, {and}
  \bibinfo{person}{David~A. Wood}.} \bibinfo{year}{2011}\natexlab{}.
\newblock \showarticletitle{The {gem5} Simulator}.
\newblock \bibinfo{journal}{\emph{{SIGARCH} Comput. Archit. News}}
  \bibinfo{volume}{39}, \bibinfo{number}{2} (\bibinfo{year}{2011}),
  \bibinfo{pages}{1--7}.
\newblock
\urldef\tempurl%
\url{https://doi.org/10.1145/2024716.2024718}
\showDOI{\tempurl}


\bibitem[B{\"{o}}hm et~al\mbox{.}(2010)]%
        {bohm10}
\bibfield{author}{\bibinfo{person}{Igor B{\"{o}}hm},
  \bibinfo{person}{Bj{\"{o}}rn Franke}, {and} \bibinfo{person}{Nigel~P.
  Topham}.} \bibinfo{year}{2010}\natexlab{}.
\newblock \showarticletitle{Cycle-Accurate Performance Modelling in an
  Ultra-Fast Just-in-Time Dynamic Binary Translation Instruction Set
  Simulator}. In \bibinfo{booktitle}{\emph{Proceedings of the 2010
  International Conference on Embedded Computer Systems: Architectures,
  Modeling and Simulation {(IC-SAMOS} 2010), Samos, Greece, July 19-22, 2010}},
  \bibfield{editor}{\bibinfo{person}{Fadi~J. Kurdahi} {and}
  \bibinfo{person}{Jarmo Takala}} (Eds.). \bibinfo{publisher}{{IEEE}},
  \bibinfo{pages}{1--10}.
\newblock
\urldef\tempurl%
\url{https://doi.org/10.1109/ICSAMOS.2010.5642102}
\showDOI{\tempurl}


\bibitem[Brubaker et~al\mbox{.}(2014)]%
        {brubaker14frankencerts}
\bibfield{author}{\bibinfo{person}{Chad Brubaker}, \bibinfo{person}{Suman
  Jana}, \bibinfo{person}{Baishakhi Ray}, \bibinfo{person}{Sarfraz Khurshid},
  {and} \bibinfo{person}{Vitaly Shmatikov}.} \bibinfo{year}{2014}\natexlab{}.
\newblock \showarticletitle{Using Frankencerts for Automated Adversarial
  Testing of Certificate Validation in {SSL/TLS} Implementations}. In
  \bibinfo{booktitle}{\emph{2014 {IEEE} Symposium on Security and Privacy, {SP}
  2014, Berkeley, CA, USA, May 18-21, 2014}}. \bibinfo{publisher}{{IEEE}
  Computer Society}, \bibinfo{pages}{114--129}.
\newblock
\urldef\tempurl%
\url{https://doi.org/10.1109/SP.2014.15}
\showDOI{\tempurl}


\bibitem[Chaitin et~al\mbox{.}(1981)]%
        {graphcol}
\bibfield{author}{\bibinfo{person}{Gregory~J. Chaitin},
  \bibinfo{person}{Marc~A. Auslander}, \bibinfo{person}{Ashok~K. Chandra},
  \bibinfo{person}{John Cocke}, \bibinfo{person}{Martin~E. Hopkins}, {and}
  \bibinfo{person}{Peter~W. Markstein}.} \bibinfo{year}{1981}\natexlab{}.
\newblock \showarticletitle{Register Allocation Via Coloring}.
\newblock \bibinfo{journal}{\emph{Comput. Lang.}} \bibinfo{volume}{6},
  \bibinfo{number}{1} (\bibinfo{year}{1981}), \bibinfo{pages}{47--57}.
\newblock
\urldef\tempurl%
\url{https://doi.org/10.1016/0096-0551(81)90048-5}
\showDOI{\tempurl}


\bibitem[Chatelet(2018)]%
        {llvmexegesis}
\bibfield{author}{\bibinfo{person}{Guillaume Chatelet}.}
  \bibinfo{year}{2018}\natexlab{}.
\newblock \bibinfo{title}{llvm-exegesis: Automatic Measurement of Instruction
  Latency/Uops}.
\newblock
\newblock
\urldef\tempurl%
\url{https://lists.llvm.org/pipermail/llvm-dev/2018-March/121814.html}
\showURL{%
\tempurl}
\newblock
\shownote{Accessed: 2021-07-22}.


\bibitem[Chen et~al\mbox{.}(2019)]%
        {chen19}
\bibfield{author}{\bibinfo{person}{Yishen Chen}, \bibinfo{person}{Ajay
  Brahmakshatriya}, \bibinfo{person}{Charith Mendis}, \bibinfo{person}{Alex
  Renda}, \bibinfo{person}{Eric Atkinson}, \bibinfo{person}{Ondrej S{\`y}kora},
  \bibinfo{person}{Saman Amarasinghe}, {and} \bibinfo{person}{Michael Carbin}.}
  \bibinfo{year}{2019}\natexlab{}.
\newblock \showarticletitle{{BHive}: A Benchmark Suite and Measurement
  Framework for Validating x86-64 Basic Block Performance Models}. In
  \bibinfo{booktitle}{\emph{2019 IEEE international symposium on workload
  characterization (IISWC). IEEE}}.
\newblock
\urldef\tempurl%
\url{https://doi.org/10.1109/IISWC47752.2019.9042166}
\showDOI{\tempurl}


\bibitem[Chen and Su(2015)]%
        {chen15mucerts}
\bibfield{author}{\bibinfo{person}{Yuting Chen} {and} \bibinfo{person}{Zhendong
  Su}.} \bibinfo{year}{2015}\natexlab{}.
\newblock \showarticletitle{Guided Differential Testing of Certificate
  Validation in {SSL/TLS} implementations}. In
  \bibinfo{booktitle}{\emph{Proceedings of the 2015 10th Joint Meeting on
  Foundations of Software Engineering, {ESEC/FSE} 2015, Bergamo, Italy, August
  30 - September 4, 2015}}, \bibfield{editor}{\bibinfo{person}{Elisabetta~Di
  Nitto}, \bibinfo{person}{Mark Harman}, {and} \bibinfo{person}{Patrick
  Heymans}} (Eds.). \bibinfo{publisher}{{ACM}}, \bibinfo{pages}{793--804}.
\newblock
\urldef\tempurl%
\url{https://doi.org/10.1145/2786805.2786835}
\showDOI{\tempurl}


\bibitem[Cousot and Cousot(1977)]%
        {cousot77}
\bibfield{author}{\bibinfo{person}{Patrick Cousot} {and}
  \bibinfo{person}{Radhia Cousot}.} \bibinfo{year}{1977}\natexlab{}.
\newblock \showarticletitle{Abstract Interpretation: {A} Unified Lattice Model
  for Static Analysis of Programs by Construction or Approximation of
  Fixpoints}. In \bibinfo{booktitle}{\emph{Conference Record of the Fourth
  {ACM} Symposium on Principles of Programming Languages, Los Angeles,
  California, USA, January 1977}}, \bibfield{editor}{\bibinfo{person}{Robert~M.
  Graham}, \bibinfo{person}{Michael~A. Harrison}, {and} \bibinfo{person}{Ravi
  Sethi}} (Eds.). \bibinfo{publisher}{{ACM}}, \bibinfo{pages}{238--252}.
\newblock
\urldef\tempurl%
\url{https://doi.org/10.1145/512950.512973}
\showDOI{\tempurl}


\bibitem[Derumigny et~al\mbox{.}(2022)]%
        {derumigny21palmed}
\bibfield{author}{\bibinfo{person}{Nicolas Derumigny},
  \bibinfo{person}{Th{\'{e}}ophile Bastian}, \bibinfo{person}{Fabian Gruber},
  \bibinfo{person}{Guillaume Iooss}, \bibinfo{person}{Christophe Guillon},
  \bibinfo{person}{Louis{-}No{\"{e}}l Pouchet}, {and} \bibinfo{person}{Fabrice
  Rastello}.} \bibinfo{year}{2022}\natexlab{}.
\newblock \showarticletitle{{PALMED:} Throughput Characterization for
  Superscalar Architectures}. In \bibinfo{booktitle}{\emph{{IEEE/ACM}
  International Symposium on Code Generation and Optimization, {CGO} 2022,
  Seoul, Korea, Republic of, April 2-6, 2022}},
  \bibfield{editor}{\bibinfo{person}{Jae~W. Lee}, \bibinfo{person}{Sebastian
  Hack}, {and} \bibinfo{person}{Tatiana Shpeisman}} (Eds.).
  \bibinfo{publisher}{{IEEE}}, \bibinfo{pages}{106--117}.
\newblock
\urldef\tempurl%
\url{https://doi.org/10.1109/CGO53902.2022.9741289}
\showDOI{\tempurl}


\bibitem[{Di Biagio}(2018)]%
        {llvmmca}
\bibfield{author}{\bibinfo{person}{Andrea {Di Biagio}}.}
  \bibinfo{year}{2018}\natexlab{}.
\newblock \bibinfo{title}{llvm-mca: A Static Performance Analysis Tool}.
\newblock
\newblock
\urldef\tempurl%
\url{http://lists.llvm.org/pipermail/llvm-dev/2018-March/121490.html}
\showURL{%
\tempurl}


\bibitem[Dutra et~al\mbox{.}(2019)]%
        {dutra19smtsampling}
\bibfield{author}{\bibinfo{person}{Rafael Dutra}, \bibinfo{person}{Jonathan
  Bachrach}, {and} \bibinfo{person}{Koushik Sen}.}
  \bibinfo{year}{2019}\natexlab{}.
\newblock \showarticletitle{{GUIDEDSAMPLER:} Coverage-guided Sampling of {SMT}
  Solutions}. In \bibinfo{booktitle}{\emph{2019 Formal Methods in Computer
  Aided Design, {FMCAD} 2019, San Jose, CA, USA, October 22-25, 2019}},
  \bibfield{editor}{\bibinfo{person}{Clark~W. Barrett} {and}
  \bibinfo{person}{Jin Yang}} (Eds.). \bibinfo{publisher}{{IEEE}},
  \bibinfo{pages}{203--211}.
\newblock
\urldef\tempurl%
\url{https://doi.org/10.23919/FMCAD.2019.8894251}
\showDOI{\tempurl}


\bibitem[Google(2018)]%
        {exegesis}
\bibfield{author}{\bibinfo{person}{Google}.} \bibinfo{year}{2018}\natexlab{}.
\newblock \bibinfo{title}{{EXEgesis}: Automatic Measurement of Instruction
  Latency/Uops}.
\newblock
\newblock
\urldef\tempurl%
\url{https://github.com/google/EXEgesis}
\showURL{%
\tempurl}
\newblock
\shownote{Accessed: 2021-07-22}.


\bibitem[Intel(2012)]%
        {iaca}
\bibfield{author}{\bibinfo{person}{Intel}.} \bibinfo{year}{2012}\natexlab{}.
\newblock \bibinfo{title}{Intel Architecture Code Analyzer}.
\newblock
\newblock
\urldef\tempurl%
\url{https://software.intel.com/en-us/articles/intel-architecture-code-analyzer}
\showURL{%
\tempurl}


\bibitem[Jay and Miller(2018)]%
        {jay2018}
\bibfield{author}{\bibinfo{person}{Nathan Jay} {and} \bibinfo{person}{Barton~P.
  Miller}.} \bibinfo{year}{2018}\natexlab{}.
\newblock \showarticletitle{Structured Random Differential Testing of
  Instruction Decoders}. In \bibinfo{booktitle}{\emph{2018 IEEE 25th
  International Conference on Software Analysis, Evolution and Reengineering
  (SANER)}}. \bibinfo{pages}{84--94}.
\newblock
\urldef\tempurl%
\url{https://doi.org/10.1109/SANER.2018.8330199}
\showDOI{\tempurl}


\bibitem[Lattner and Adve(2004)]%
        {lattner04}
\bibfield{author}{\bibinfo{person}{Chris Lattner} {and} \bibinfo{person}{Vikram
  Adve}.} \bibinfo{year}{2004}\natexlab{}.
\newblock \showarticletitle{{{LLVM}: A Compilation Framework for Lifelong
  Program Analysis \& Transformation}}. In
  \bibinfo{booktitle}{\emph{{Proceedings of the 2004 International Symposium on
  Code Generation and Optimization (CGO'04)}}}. \bibinfo{address}{Palo Alto,
  California}.
\newblock
\urldef\tempurl%
\url{https://doi.org/10.1109/CGO.2004.1281665}
\showDOI{\tempurl}


\bibitem[Laukemann et~al\mbox{.}(2018)]%
        {laukemann18osaca}
\bibfield{author}{\bibinfo{person}{Jan Laukemann}, \bibinfo{person}{Julian
  Hammer}, \bibinfo{person}{Johannes Hofmann}, \bibinfo{person}{Georg Hager},
  {and} \bibinfo{person}{Gerhard Wellein}.} \bibinfo{year}{2018}\natexlab{}.
\newblock \showarticletitle{Automated Instruction Stream Throughput Prediction
  for Intel and AMD Microarchitectures}. In \bibinfo{booktitle}{\emph{2018
  IEEE/ACM Performance Modeling, Benchmarking and Simulation of High
  Performance Computer Systems (PMBS)}}. IEEE, \bibinfo{pages}{121--131}.
\newblock
\urldef\tempurl%
\url{https://doi.org/10.1109/PMBS.2018.8641578}
\showDOI{\tempurl}


\bibitem[{LLVM}(2021)]%
        {llvm-zen-model}
\bibfield{author}{\bibinfo{person}{{LLVM}}.} \bibinfo{year}{2021}\natexlab{}.
\newblock \bibinfo{title}{{LLVM} 13 Scheduling Model for AMD Zen/Zen+ CPUs}.
\newblock
  \bibinfo{howpublished}{\url{https://github.com/llvm/llvm-project/blob/release/13.x/llvm/lib/Target/X86/X86ScheduleZnver1.td}}.
\newblock
\newblock
\shownote{Accessed: 2022-04-07}.


\bibitem[LLVM(2021a)]%
        {llvmexegesisman}
\bibfield{author}{\bibinfo{person}{LLVM}.} \bibinfo{year}{2021}\natexlab{a}.
\newblock \bibinfo{title}{llvm-exegesis - LLVM Machine Instruction Benchmark}.
\newblock
\newblock
\urldef\tempurl%
\url{https://llvm.org/docs/CommandGuide/llvm-exegesis.html}
\showURL{%
\tempurl}
\newblock
\shownote{Accessed: 2021-07-22}.


\bibitem[LLVM(2021b)]%
        {llvmmca-man}
\bibfield{author}{\bibinfo{person}{LLVM}.} \bibinfo{year}{2021}\natexlab{b}.
\newblock \bibinfo{title}{llvm-mca - LLVM Machine Code Analyzer}.
\newblock
\newblock
\urldef\tempurl%
\url{https://llvm.org/docs/CommandGuide/llvm-mca.html}
\showURL{%
\tempurl}
\newblock
\shownote{Accessed: 2021-11-15}.


\bibitem[Manès et~al\mbox{.}(2019)]%
        {manes19fuzzsurvey}
\bibfield{author}{\bibinfo{person}{Valentin Jean~Marie Manès},
  \bibinfo{person}{HyungSeok Han}, \bibinfo{person}{Choongwoo Han},
  \bibinfo{person}{Sang~Kil Cha}, \bibinfo{person}{Manuel Egele},
  \bibinfo{person}{Edward~J. Schwartz}, {and} \bibinfo{person}{Maverick Woo}.}
  \bibinfo{year}{2019}\natexlab{}.
\newblock \showarticletitle{The Art, Science, and Engineering of Fuzzing: A
  Survey}.
\newblock \bibinfo{journal}{\emph{IEEE Transactions on Software Engineering}}
  (\bibinfo{year}{2019}), \bibinfo{pages}{1--1}.
\newblock
\urldef\tempurl%
\url{https://doi.org/10.1109/TSE.2019.2946563}
\showDOI{\tempurl}


\bibitem[McKeeman(1998)]%
        {mckeeman98}
\bibfield{author}{\bibinfo{person}{William~M. McKeeman}.}
  \bibinfo{year}{1998}\natexlab{}.
\newblock \showarticletitle{Differential Testing for Software}.
\newblock \bibinfo{journal}{\emph{Digit. Tech. J.}} \bibinfo{volume}{10},
  \bibinfo{number}{1} (\bibinfo{year}{1998}), \bibinfo{pages}{100--107}.
\newblock
\urldef\tempurl%
\url{http://www.hpl.hp.com/hpjournal/dtj/vol10num1/vol10num1art9.pdf}
\showURL{%
\tempurl}


\bibitem[Mendis et~al\mbox{.}(2019)]%
        {mendis19ithemal}
\bibfield{author}{\bibinfo{person}{Charith Mendis}, \bibinfo{person}{Alex
  Renda}, \bibinfo{person}{Saman Amarasinghe}, {and} \bibinfo{person}{Michael
  Carbin}.} \bibinfo{year}{2019}\natexlab{}.
\newblock \showarticletitle{Ithemal: Accurate, Portable and Fast Basic Block
  Throughput Estimation using Deep Neural Networks}. In
  \bibinfo{booktitle}{\emph{Proceedings of the 36th International Conference on
  Machine Learning}} \emph{(\bibinfo{series}{Proceedings of Machine Learning
  Research}, Vol.~\bibinfo{volume}{97})},
  \bibfield{editor}{\bibinfo{person}{Kamalika Chaudhuri} {and}
  \bibinfo{person}{Ruslan Salakhutdinov}} (Eds.). \bibinfo{publisher}{PMLR},
  \bibinfo{address}{Long Beach, California, USA}, \bibinfo{pages}{4505--4515}.
\newblock
\urldef\tempurl%
\url{http://proceedings.mlr.press/v97/mendis19a.html}
\showURL{%
\tempurl}


\bibitem[Ofenbeck et~al\mbox{.}(2014)]%
        {ofenbeck14roofline}
\bibfield{author}{\bibinfo{person}{Georg Ofenbeck}, \bibinfo{person}{Ruedi
  Steinmann}, \bibinfo{person}{Victoria~Caparr{\'{o}}s Cabezas},
  \bibinfo{person}{Daniele~G. Spampinato}, {and} \bibinfo{person}{Markus
  P{\"{u}}schel}.} \bibinfo{year}{2014}\natexlab{}.
\newblock \showarticletitle{Applying the Roofline Model}. In
  \bibinfo{booktitle}{\emph{2014 {IEEE} International Symposium on Performance
  Analysis of Systems and Software, {ISPASS} 2014, Monterey, CA, USA, March
  23-25, 2014}}. \bibinfo{publisher}{{IEEE} Computer Society},
  \bibinfo{pages}{76--85}.
\newblock
\urldef\tempurl%
\url{https://doi.org/10.1109/ISPASS.2014.6844463}
\showDOI{\tempurl}


\bibitem[Oleksenko et~al\mbox{.}(2021)]%
        {oleksenko21revizor}
\bibfield{author}{\bibinfo{person}{Oleksii Oleksenko},
  \bibinfo{person}{Christof Fetzer}, \bibinfo{person}{Boris K{\"{o}}pf}, {and}
  \bibinfo{person}{Mark Silberstein}.} \bibinfo{year}{2021}\natexlab{}.
\newblock \showarticletitle{Revizor: Fuzzing for Leaks in Black-box CPUs}.
\newblock \bibinfo{journal}{\emph{CoRR}}  \bibinfo{volume}{abs/2105.06872}
  (\bibinfo{year}{2021}).
\newblock
\showeprint[arXiv]{2105.06872}
\urldef\tempurl%
\url{https://arxiv.org/abs/2105.06872}
\showURL{%
\tempurl}


\bibitem[Paleari et~al\mbox{.}(2010)]%
        {paleari2010}
\bibfield{author}{\bibinfo{person}{Roberto Paleari}, \bibinfo{person}{Lorenzo
  Martignoni}, \bibinfo{person}{Giampaolo Fresi~Roglia}, {and}
  \bibinfo{person}{Danilo Bruschi}.} \bibinfo{year}{2010}\natexlab{}.
\newblock \showarticletitle{N-Version Disassembly: Differential Testing of X86
  Disassemblers}. In \bibinfo{booktitle}{\emph{Proceedings of the 19th
  International Symposium on Software Testing and Analysis}} (Trento, Italy)
  \emph{(\bibinfo{series}{ISSTA '10})}. \bibinfo{publisher}{Association for
  Computing Machinery}, \bibinfo{address}{New York, NY, USA},
  \bibinfo{pages}{265–274}.
\newblock
\showISBNx{9781605588230}
\urldef\tempurl%
\url{https://doi.org/10.1145/1831708.1831741}
\showDOI{\tempurl}


\bibitem[Parnin and Orso(2011)]%
        {parnin2011}
\bibfield{author}{\bibinfo{person}{Chris Parnin} {and}
  \bibinfo{person}{Alessandro Orso}.} \bibinfo{year}{2011}\natexlab{}.
\newblock \showarticletitle{Are Automated Debugging Techniques Actually Helping
  Programmers?}. In \bibinfo{booktitle}{\emph{Proceedings of the 20th
  International Symposium on Software Testing and Analysis, {ISSTA} 2011,
  Toronto, ON, Canada, July 17-21, 2011}},
  \bibfield{editor}{\bibinfo{person}{Matthew~B. Dwyer} {and}
  \bibinfo{person}{Frank Tip}} (Eds.). \bibinfo{publisher}{{ACM}},
  \bibinfo{pages}{199--209}.
\newblock
\urldef\tempurl%
\url{https://doi.org/10.1145/2001420.2001445}
\showDOI{\tempurl}


\bibitem[Petsios et~al\mbox{.}(2017)]%
        {petsios17nezha}
\bibfield{author}{\bibinfo{person}{Theofilos Petsios}, \bibinfo{person}{Adrian
  Tang}, \bibinfo{person}{Salvatore Stolfo}, \bibinfo{person}{Angelos~D.
  Keromytis}, {and} \bibinfo{person}{Suman Jana}.}
  \bibinfo{year}{2017}\natexlab{}.
\newblock \showarticletitle{{NEZHA}: Efficient Domain-Independent Differential
  Testing}. In \bibinfo{booktitle}{\emph{2017 IEEE Symposium on Security and
  Privacy (SP)}}. \bibinfo{pages}{615--632}.
\newblock
\urldef\tempurl%
\url{https://doi.org/10.1109/SP.2017.27}
\showDOI{\tempurl}


\bibitem[Renda et~al\mbox{.}(2020)]%
        {renda2020difftune}
\bibfield{author}{\bibinfo{person}{Alex Renda}, \bibinfo{person}{Yishen Chen},
  \bibinfo{person}{Charith Mendis}, {and} \bibinfo{person}{Michael Carbin}.}
  \bibinfo{year}{2020}\natexlab{}.
\newblock \showarticletitle{{DiffTune}: Optimizing {CPU} Simulator Parameters
  with Learned Differentiable Surrogates}. In \bibinfo{booktitle}{\emph{2020
  53rd Annual IEEE/ACM International Symposium on Microarchitecture (MICRO)}}.
  IEEE, \bibinfo{pages}{442--455}.
\newblock
\urldef\tempurl%
\url{https://doi.org/10.1109/MICRO50266.2020.00045}
\showDOI{\tempurl}


\bibitem[Ritter and Hack(2020)]%
        {pmevo}
\bibfield{author}{\bibinfo{person}{Fabian Ritter} {and}
  \bibinfo{person}{Sebastian Hack}.} \bibinfo{year}{2020}\natexlab{}.
\newblock \showarticletitle{PMEvo: Portable Inference of Port Mappings for
  Out-of-Order Processors by Evolutionary Optimization}. In
  \bibinfo{booktitle}{\emph{Proceedings of the 41st {ACM} {SIGPLAN}
  International Conference on Programming Language Design and Implementation,
  {PLDI} 2020, London, UK, June 15-20, 2020}},
  \bibfield{editor}{\bibinfo{person}{Alastair~F. Donaldson} {and}
  \bibinfo{person}{Emina Torlak}} (Eds.). \bibinfo{publisher}{{ACM}},
  \bibinfo{pages}{608--622}.
\newblock
\urldef\tempurl%
\url{https://doi.org/10.1145/3385412.3385995}
\showDOI{\tempurl}


\bibitem[Ritter and Hack(2022)]%
        {anica-artifact}
\bibfield{author}{\bibinfo{person}{Fabian Ritter} {and}
  \bibinfo{person}{Sebastian Hack}.} \bibinfo{year}{2022}\natexlab{}.
\newblock \bibinfo{booktitle}{\emph{{AnICA: Analyzing Inconsistencies in
  Microarchitectural Code Analyzers (Artifact)}}}.
\newblock
\urldef\tempurl%
\url{https://doi.org/10.5281/zenodo.6818171}
\showDOI{\tempurl}


\bibitem[Rubial et~al\mbox{.}(2014)]%
        {rubial14cqa}
\bibfield{author}{\bibinfo{person}{Andres~Charif Rubial},
  \bibinfo{person}{Emmanuel Oseret}, \bibinfo{person}{Jose Noudohouenou},
  \bibinfo{person}{William Jalby}, {and} \bibinfo{person}{Ghislain Lartigue}.}
  \bibinfo{year}{2014}\natexlab{}.
\newblock \showarticletitle{{CQA:} {A} Code Quality Analyzer Tool at Binary
  Level}. In \bibinfo{booktitle}{\emph{21st International Conference on High
  Performance Computing, HiPC 2014, Goa, India, December 17-20, 2014}}.
  \bibinfo{publisher}{{IEEE} Computer Society}, \bibinfo{pages}{1--10}.
\newblock
\urldef\tempurl%
\url{https://doi.org/10.1109/HiPC.2014.7116904}
\showDOI{\tempurl}


\bibitem[Williams et~al\mbox{.}(2009)]%
        {williams09roofline}
\bibfield{author}{\bibinfo{person}{Samuel Williams}, \bibinfo{person}{Andrew
  Waterman}, {and} \bibinfo{person}{David~A. Patterson}.}
  \bibinfo{year}{2009}\natexlab{}.
\newblock \showarticletitle{Roofline: An Insightful Visual Performance Model
  for Multicore Architectures}.
\newblock \bibinfo{journal}{\emph{Commun. {ACM}}} \bibinfo{volume}{52},
  \bibinfo{number}{4} (\bibinfo{year}{2009}), \bibinfo{pages}{65--76}.
\newblock
\urldef\tempurl%
\url{https://doi.org/10.1145/1498765.1498785}
\showDOI{\tempurl}


\bibitem[Woodruff et~al\mbox{.}(2021)]%
        {woodruff2021differential}
\bibfield{author}{\bibinfo{person}{William Woodruff}, \bibinfo{person}{Niki
  Carroll}, {and} \bibinfo{person}{Sebastiaan Peters}.}
  \bibinfo{year}{2021}\natexlab{}.
\newblock \bibinfo{booktitle}{\emph{Differential Analysis of x86-64 Instruction
  Decoders}}.
\newblock \bibinfo{type}{{T}echnical {R}eport}. \bibinfo{pages}{152--161}
  pages.
\newblock
\urldef\tempurl%
\url{https://doi.org/10.1109/SPW53761.2021.00029}
\showDOI{\tempurl}


\end{thebibliography}
